\documentclass[14pt]{article} 
\usepackage[sectionbib]{natbib}
\usepackage{array,epsfig,fancyheadings,rotating}

\usepackage{amsmath}
\usepackage{amssymb}
\usepackage{amsfonts}
\usepackage{amsthm}
\usepackage{graphicx}
\usepackage{color}
\usepackage{comment}
 \usepackage{listings}

 \usepackage{url}
 
\newtheorem{theorem}{Theorem}
\newtheorem{lemma}{Lemma}
\newtheorem{corollary}{Corollary}
\newtheorem{proposition}{Proposition}
\theoremstyle{definition}
\newtheorem{remark}{Remark}

\newcommand{\bA}{\mbox{\boldmath {$A$}}}

\newcommand{\bB}{\mbox{\boldmath {$B$}}}

\newcommand{\bh}{\mbox{\boldmath {$h$}}}
\newcommand{\bH}{\mbox{\boldmath {$H$}}}

\newcommand{\bI}{\mbox{\boldmath {$I$}}}

\newcommand{\bM}{\mbox{\boldmath {$M$}}}

\newcommand{\bO}{\mbox{\boldmath {$O$}}}

\newcommand{\bP}{\mbox{\boldmath {$P$}}}

\newcommand{\bS}{\mbox{\boldmath {$S$}}}

\newcommand{\bu}{\mbox{\boldmath {$u$}}}

\newcommand{\bv}{\mbox{\boldmath {$v$}}}

\newcommand{\bx}{\mbox{\boldmath {$x$}}}
\newcommand{\bX}{\mbox{\boldmath {$X$}}}

\newcommand{\bz}{\mbox{\boldmath {$z$}}}
\newcommand{\bZ}{\mbox{\boldmath {$Z$}}}
\newcommand{\bze}{\mbox{\boldmath {$0$}}}
\newcommand{\bone}{\mbox{\boldmath {$1$}}}

\newcommand{\bmu}{\mbox{\boldmath $ \mu $}}
\newcommand{\bSig}{\mbox{\boldmath $ \Sigma $}}

\newcommand{\bSigma}{\mbox{\boldmath $ \Sigma $}}

\newcommand{\bLam}{\mbox{\boldmath $ \Lambda $}}
\newcommand{\bbeta}{\mbox{\boldmath $ \beta $}}

\newcommand{\bOme}{\mbox{\boldmath $ \Omega $}}
\newcommand{\bGamma}{\mbox{\boldmath $\Gamma$}}
\newcommand{\bPsi}{\mbox{\boldmath $\Psi$}}

\newcommand{\bep}{\mbox{\boldmath $ \varepsilon $}}

\newcommand{\tr}{\mbox{tr}}
\newcommand{\argmin}{\mathop{\rm argmin}\limits}

\newcommand{\Var}{\mbox{Var}}
\newcommand\CG[1]{\textcolor{black}{#1}}

\newcommand{\bsB}{\mbox{\scriptsize $\bB$}}

\begin{document}


$\ $\par
\vspace{6mm}


\centerline{\large\bf 
Automatic sparse PCA for high-dimensional data}
\vspace{.4cm} 
\centerline{Kazuyoshi Yata and Makoto Aoshima} 
\vspace{.4cm} 
\centerline{\it Institute of Mathematics, University of Tsukuba}
 \vspace{.55cm}


\begin{quotation}
Sparse principal component analysis (SPCA) methods have proven to efficiently analyze high-dimensional data. 
Among them, threshold-based SPCA (TSPCA) is computationally more cost-effective than regularized SPCA, based on L1 penalties.
We herein present an investigation of the efficacy of TSPCA for high-dimensional data settings and illustrate that, for a suitable threshold value, 
TSPCA achieves satisfactory performance for high-dimensional data.
Thus, the performance of the TSPCA depends heavily on the selected threshold value. 
To this end, we propose a novel thresholding estimator to obtain the principal component (PC) directions using a customized noise-reduction methodology. 
The proposed technique is consistent under mild conditions, unaffected by threshold values, and therefore yields more accurate results quickly at a lower computational cost.
Furthermore, we explore the shrinkage PC directions and their application in clustering high-dimensional data. Finally, we evaluate the performance of the estimated shrinkage PC directions in actual data analyses.

\vspace{9pt}
\par
\end{quotation}\par

\def\thefigure{\arabic{figure}}
\def\thetable{\arabic{table}}
\renewcommand{\theequation}{\thesection.\arabic{equation}}

\fontsize{12}{14pt plus.8pt minus .6pt}\selectfont

\section{Introduction}
\label{sec:intro}
\setcounter{equation}{0}

High-dimensional, low-sample-size (HDLSS) data scenarios exist in many areas of modern
science including genomics, medical imaging, text recognition, and finance. 
In recent years, substantial work has been conducted on HDLSS asymptotic theory, 
wherein the sample size $n$ is fixed or $n/d\to 0$ is used as the data dimension $d\to\infty$. 
For principal component analysis (PCA), 
\cite{Jung:2009} and \cite{Yata:2009} investigated inconsistency properties for both the eigenvalues and principal component (PC) 
directions in a sample covariance matrix. 
\cite{Yata:2012} developed a new PCA method called the {\it noise-reduction methodology} and reported consistent estimators for both eigenvalues and PC directions in addition with the PC scores using this method. 
Sparse PCA (SPCA) methods have been investigated in several studies. For example, 
\cite{Zou:2006}, \cite{Shen:2008}, and \cite{Lee:2010b} considered a regularized SPCA (RSPCA) based on L1 penalties 
under high-dimensional settings.
\cite{Johnstone:2009} proposed a thresholded SPCA (TSPCA) and presented a consistency property of the TSPCA when $n/d\to 0$.  
Further, \cite{Shen:2013} showed that the PC directions obtained by RSPCA and TSPCA are consistent when $d\to \infty$ while $n$ is fixed. 
In addition, \cite{Paul:2007b} developed an augmented SPCA method and \cite{Ma:2013} proposed an iterative thresholding procedure for PC directions. 
In this study, we focused on TSPCA under high-dimensional settings.

Suppose that we have a $d\times n$ data matrix $\bX=(\bx_{1},...,\bx_{n})$, where $\bx_{i}=(x_{i(1)},...,x_{i(d)})^T$, $i= 1,...,n$, are independent and identically distributed (i.i.d.) as a $d$-dimensional distribution with 
mean $\bmu$ and (non-negative definite) covariance matrix $\bSig$. 
We express 
the eigen-decomposition of $\bSig$ as $\bSig=\bH \bLam \bH^T$, where $\bLam$ represents a diagonal matrix of the eigenvalues, 
$\lambda_{1}\ge \cdots \ge \lambda_{d}(\ge 0)$, and 
$\bH=(\bh_{1},...,\bh_{d})$ is an orthogonal matrix of the corresponding eigenvectors. 
The sample covariance matrix is given by 
$\bS=(n-1)^{-1}(\bX-\overline{\bX})(\bX-\overline{\bX})^T=(n-1)^{-1}\sum_{i=1}^n(\bx_i-\bar{\bx})(\bx_i-\bar{\bx})^T$, where $\bar{\bx}=n^{-1}\sum_{i=1}^n \bx_i$ and $\overline{\bX}=\bar{\bx}\bone_n^T$ with $\bone_n=(1,...,1)^T$. 
Let $\hat{\lambda}_1\ge\cdots\ge\hat{\lambda}_d \ge 0$ be the eigenvalues of $\bS$, and 
let $\hat{\bh}_j,\ j=1,...,d$ be the corresponding eigenvectors $\hat{\bh}_j^T\hat{\bh}_{j'}=\delta_{jj'}$, 
\CG{
where $\delta_{jj'}$ is the Kronecker delta.}  
Thus, the eigen-decomposition of $\bS$ is 
$
\bS=\sum_{s=1}^d \hat{\lambda}_s\hat{\bh}_s\hat{\bh}_s^T.
$ 
We assume that $\bh_{j}^T\hat{\bh}_{j} \ge 0$ for all $j$ without the loss of generality. 
We now consider the $n \times n$ dual-sample covariance matrix defined by $\bS_D=(n-1)^{-1}(\bX-\overline{\bX})^T(\bX-\overline{\bX})$. 
Here, $\bS$ and $\bS_D$ share nonzero eigenvalues. 
Let the eigen-decomposition of $\bS_{D}$ be $\bS_{D}=\sum_{j=1}^{n-1}\hat{\lambda}_{j}\hat{\bu}_{j}\hat{\bu}_{j}^T $, 
where $\hat{\bu}_{j}$ denotes an eigenvector corresponding to $\hat{\lambda}_{j}$ and $\hat{\bu}_{j}^T\hat{\bu}_{j'}=\delta_{jj'}$. 
Furthermore, $\hat{\bh}_{j}$ can be calculated as
$
\hat{\bh}_{j}=\{(n-1)\hat{\lambda}_{j}\}^{-1/2}(\bX-\overline{\bX}) \hat{\bu}_{j}.
$

\cite{Johnstone:2001}, \cite{Baik:2006}, and \cite{Paul:2007} considered a spiked model for the eigenvalues.  
\begin{align}
&\mbox{$\lambda_j\ (>\kappa)$, $j=1,...,m,$ are fixed (not depending on $d$)} \notag \\
&\mbox{and } \lambda_{m+1}=\cdots =\lambda_d=\kappa.
\label{1.1}
\end{align}
Here, $m$ represents a fixed positive integer and $\kappa \ (>0)$ represents a fixed constant. 
Under (\ref{1.1}), 
the asymptotic behavior of the eigenvalues of $\bS$ was studied when both $d$ and $n$ increased at the same rate, that is, from $n/d\to \gamma>0$. 
Details under the Gaussian assumptions were reported by \cite{Johnstone:2001}, \cite{Johnstone:2009}, and \cite{Paul:2007}. 
Further, \cite{Baik:2006} and \cite{Lee:2010} reported details under the non-Gaussian but i.i.d. assumptions as in (\ref{A'}). 
For review, the authors direct readers to \cite{Paul:2014}.
The condition $\lambda_{m+1}=\cdots =\lambda_d=\kappa$ is strict for the latter part of (\ref{1.1}).  
Without assuming that $\lambda_{m+1}=\cdots =\lambda_d=\kappa$ in (\ref{1.1}), \cite{Bai:2012} estimated the forward eigenvalues. 
However, the former part of (\ref{1.1}) is a strict condition because the eigenvalues depend on $d$, and it is probable that $\lambda_j\to\infty$ for the first several $j$s when $d\to \infty$. 
Details were provided by \cite{Fan:2013}, \cite{Jung:2009}, \cite{Onatski:2012}, \cite{Shen:2016}, \cite{Wang:2017}, and \cite{Yata:2012,Yata:2013}. 
They considered spiked models such as  
\begin{equation}
\lambda_j=\kappa_j d^{\alpha_j} \ (j=1,...,m) \quad \mbox{and} \quad \lambda_j=\kappa_j \ (j=m+1,...,d).
\label{spike}
\end{equation}
Here, $\kappa_j\ (> 0)$ and $\alpha_j\ (\alpha_1\ge\cdots\ge\alpha_m>0)$ are fixed constants 
preserving the order that $\lambda_1\ge\cdots\ge\lambda_d$. 
For example,
\cite{Cai:2020}, \cite{Shen:2016}, \cite{Wang:2017}, and \cite{Yata:2012} showed that 
\begin{align}
&\hat{\lambda}_j/\lambda_{j}=1+\delta/\lambda_j+o_P(1) \ 
\mbox{ as $d\to \infty$ and $n\to \infty$ for $j=1,...,m$} \notag 
\end{align}
under (\ref{spike}), $d^{1-2\alpha_m}/n=o(1)$, and (\ref{A'}), where 
$\delta={\sum_{s=m+1}^d\lambda_s}/{(n-1)}$. 
Further details are provided in Appendix E in the online supplementary material. 
Here, 
$\delta=O(d/n)$ for (\ref{spike}); if  $\delta/\lambda_j \to \infty$, $\hat{\lambda}_{j}$ is strongly inconsistent in the sense that $\lambda_{j}/\hat{\lambda}_{j}=o_P(1)$.  
\cite{Jung:2009} and  
\cite{Yata:2009} 
have reported on the concept of strong inconsistency. 
\cite{Yata:2012} proposed a noise-reduction (NR)
methodology that uses the geometric representation of 
high-dimensional eigenspaces to overcome the curse of dimensionality. 
If the NR method is used, $\lambda_{j}$s can be estimated by
\begin{equation}
\tilde{\lambda}_{j}=\hat{\lambda}_{j}-\frac{\tr(\bS_D)-\sum_{s=1}^{j}\hat{\lambda}_{s} }{n-j-1}\quad (j=1,...,n-1).
\label{NR}
\end{equation}
\cite{Yata:2012,Yata:2013} showed the consistency as ``$\tilde{\lambda}_{j}/\lambda_j=1+o_P(1)$" even when $\delta/\lambda_j \to \infty$. 
Section 2.3 provides further details. 
For PC direction $\hat{\bh}_{j}$, 
under (\ref{spike}) and some regularity conditions, 
\cite{Yata:2012,Yata:2013} and \cite{Shen:2016} showed that 
$\mbox{Angle}(\hat{\bh}_j,\bh_j)=\mbox{Arccos}\{ (1+{\delta}/{\lambda_j})^{-1/2}\}+o_P(1)$ as $d\to \infty$ and $n\to \infty$ for $j=1,...,m$.
If $\delta/\lambda_j \to \infty$, $\hat{\bh}_j$ is strongly inconsistent in that 
$\mbox{Angle}(\hat{\bh}_j,\bh_j)=\pi/2+o_P(1)$. 
To overcome this inconvenience, 
\cite{Shen:2013} showed that estimators of $\bh_1$ given by TSPCA and RSPCA  have a consistency property 
and that
they perform equivalently for high-dimensional data. 
However, 
TSPCA is easier to handle as compared to RSPCA. 
\CG{ Appendix C in the online supplementary material provides further details in this regard}. 
TSPCA can be summarized as follows: 
Let $\hat{\bh}_j=(\hat{h}_{j(1)},...,\hat{h}_{j(d)})^T$ for all $j$. 
Given a sequence of threshold values $\zeta>0$, we can define the thresholded entries as
\begin{align}
\hat{h}_{j*(j')}=
\begin{cases}
\hat{h}_{j(j')} & \mbox{if $|\hat{h}_{j(j')}|\ge\min\{\zeta, \hat{h}_{j \max} \}$}, \\
0 & \mbox{otherwise} \label{ST1}
\end{cases} 
 \quad \mbox{ for $j'=1,...,d$},
\end{align}
where $\hat{h}_{j \max} =\max_{s=1,...,d}|\hat{h}_{j(s)}|$. 
Let $\hat{\bh}_{j*}=(\hat{h}_{j*(1)},...,\hat{h}_{j*(d)})^T$. 
Then, the thresholded estimator of $\bh_j$ is defined as
\begin{align}
\hat{\bh}_{j(\zeta)}
=\hat{\bh}_{j*}/\|\hat{\bh}_{j*} \|,
 \label{ST2}
\end{align}
where $\| \cdot \|$ denotes the Euclidean norm. 
\cite{Shen:2013} showed the consistency property 
\begin{align}
\mbox{Angle}(\hat{\bh}_{1(\zeta)},\bh_1)=o_P(1)  
\mbox{ as $d\to \infty$ for some $\zeta$} \notag
\end{align}
under (\ref{spike}), 
the Gaussian assumption, and some regularity conditions.
Further, they showed consistency even when  \CG{$\delta/\lambda_j \to \infty$}. 
However, this estimator depends heavily on the choice of $\zeta$. 
  
\begin{figure}[h]
\begin{center}
\includegraphics[width=136mm]{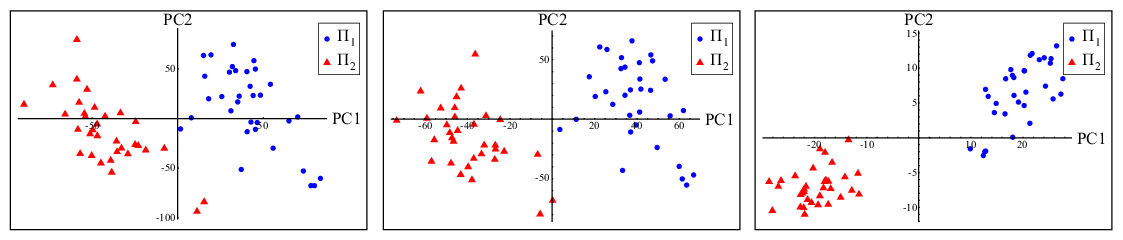} \\[-3mm]
{\footnotesize  
\ (i) Conventional PCA \ \hspace{10mm} (ii) TSPCA with $\zeta=0.01$ \ \ (iii) TSPCA with $\zeta=0.05$.}
\caption{ 
Scatter plots of the first two PC scores for 
(i) Conventional PCA, 
(ii) TSPCA with $\zeta=0.01$, and (iii) TSPCA with $\zeta=0.05$. 
 \label{F1}}
\end{center}
\end{figure}

We analyzed the microarray data given by \cite{Chiaretti:2004} wherein the dataset comprises $12625\ (=d)$ genes. 
The dataset consists of 
two tumor cellular subtypes: B-cell (33 samples) and T-cell (33 samples). 
B-cell originally contained 95 samples. 
We used only the first $33$ samples to maintain balance in the sample sizes with the T-cell. 
For the $66\ (=n)$ samples, the scatter plots of the first two PC scores are shown in Fig. \ref{F1}
for three PCAs:  
(i) Conventional PCA, 
(ii) TSPCA with $\zeta=0.01$, and (iii) TSPCA with $\zeta=0.05$. 
We have that 
$\pi/2-$Angle$(\hat{\bh}_{1(0.01)},\hat{\bh}_{2(0.01)})=0.154$, and 
$\pi/2-$Angle$(\hat{\bh}_{1(0.05)},\hat{\bh}_{2(0.05)})=0.266$. 
Further, we observed that the conventional PCA effectively classifies 
the dataset into two groups using the first PC score. 
The theoretical clarification was provided by \cite{Yata:2019}. 
The TSPCA with $\zeta=0.05$ separates them more clearly than that when using the conventional PCA. 
However, this may not hold consistency as ``$\mbox{Angle}\big(\hat{\bh}_{j(\zeta)},\bh_j\big)=o_P(1)$ for $j=1,2$" because $\pi/2-$Angle$(\hat{\bh}_{1(0.05)},\hat{\bh}_{2(0.05)})=0.266$. 
Thus, TSPCA provides preferable performance if one chooses a suitable $\zeta$, 
while it 
may be inconsistent. 

In this study, we investigated TSPCA under high-dimensional settings. 
The contributions of this study are as follows.
(I) We propose a new thresholding estimator for PC directions and present its
consistency property that holds freely from 
threshold values. 
(II) We propose a shrinkage PC direction and apply it to clustering. 

The remainder of this paper is organized as follows. 
In Section 2, we present the asymptotic properties of the NR method under certain conditions.
In Section 3, we modify the estimator of the PC directions derived using the NR method and propose a new TSPCA method. 
In Section 4, we investigate the performance of the proposed TSPCA through simulations. 
In Section 5, we present the estimation of the shrinkage PC directions and their application to clustering.  
In Section 6, we investigate the performance of the estimated shrinkage PC directions in actual data analyses. 
In the online supplementary material, we apply the proposed TSPCA to the estimation of the intrinsic component of $\bSig$. 
\section{Preliminary}
\setcounter{equation}{0} 

In this section, we lay out basic conditions, assumptions, and asymptotics for the construction of our PCA methods. 
\subsection{Strongly spiked eigenstructures}
\cite{Aoshima:2018b} 
provided two disjointed high-dimensional models:
the strongly spiked eigenvalue (SSE) model defined as 
\begin{equation}
\liminf_{d\to \infty}
\frac{\lambda_{1}^2}{\tr(\bSigma^2)}
>0 
\label{SSE}
\end{equation}
and the non-SSE (NSSE) model defined by 
\begin{equation}
 \frac{\lambda_{1}^2}{\tr(\bSigma^2)}\to 0 \quad \mbox{as $d\to \infty$}.
\label{NSSE}
\end{equation}
Notably, (\ref{NSSE}) is equivalent to ``$\tr(\bSig^4)/\tr(\bSig^2)^2\to0$ as $d\to \infty$''.
If $\alpha_1\ge 0.5$ in (\ref{spike}), then the SSE model (\ref{SSE}) holds. 
In contrast, if $\alpha_1< 0.5$ in (\ref{spike}), the NSSE model (\ref{NSSE}) holds. 
We provide additional examples of the SSE model in Remark \ref{Rem3} in Section 5.2 and Appendix D in the online supplementary material. 
The two models are essential for statistical inference of high-dimensional data. 
We emphasize that it is not possible to ensure the accuracy of high-dimensional statistical inferences using the SSE model. 
The work by  \cite{Aoshima:2018b} can be referred to for further details. 
\cite{Aoshima:2018b,Aoshima:2018a} proposed data-transformation methods based on the strongly spiked eigenstructures to overcome this inconvenience. 
Further, \cite{Yata:2019} considered clustering under the SSE model. 
The key to these references is the estimation of the strongly spiked eigenstructures. 
In this study, we focused on estimating the strongly spiked eigenstructures. 

Let $\bSig = \bSig_{1}+\bSig_2 $, where $\bSig_{1}=\sum_{s=1}^m \lambda_s\bh_s\bh_s^T $ and 
$\bSig_{2}=\sum_{s=m+1}^d \lambda_s\bh_s\bh_s^T$ 
with an unknown and positive fixed integer $m$ (independent of $d$). 
Here, $\bSig_{1}$ is considered an intrinsic part and $\bSig_{2}$, a noise component. 
We assume the following model:
\begin{description}
\item[(C-i)] \ $\lambda_{1},...,\lambda_{m}$ are distinct in that 
$
\liminf_{d\to \infty}(\lambda_{j}/\lambda_{j'}-1)>0
$ 
for $1\le j<j' \le m$ when $m\ge 2$ and $\lambda_{m}$ and $\lambda_{m+1}$ satisfy
$$
\liminf_{d\to \infty}\frac{\lambda_{m}^2}{\tr(\bSig_{2}^2) }>0 \ \mbox{ and } \ 
\frac{\lambda_{m+1}^2}{\tr(\bSig_{2}^2) }\to 0 \ \mbox{ as $d\to \infty$}.
$$ 
\end{description}
\begin{remark}
\CG{
(C-i) is an SSE model. 
The spiked model (\ref{spike}) with $\alpha_{m}\ge 0.5$ and $\kappa_j \neq \kappa_{j'}$ for $1\le j\neq j' \le m$ satisfies (C-i). 
\cite{Aoshima:2018b} provided a method to check whether the SSE model holds or not.}  
\cite{Aoshima:2018b} also provided a consistent estimator of $m$ in (C-i). 
\end{remark}

Next, we consider a bounded condition for diagonal elements. 
Let $\sigma_{(j)}=\Var(x_{i(j)})$ for all $j$. 
Here, $\sigma_{(j)}$ represents the $j$-th diagonal element of $\bSig$. 
Let $\bA_{1}=\sum_{s=1}^m\bh_{s}\bh_{s}^T$ and $\bA_{2}=\bI_d-\bA_{1}=\sum_{s=m+1}^d\bh_{s}\bh_{s}^T$, where 
$\bI_d$ represents the $d$-dimensional identity matrix. 
Let  
$\bx_{i,1}=(x_{i(1),1},...,x_{i(d),1})^T=\bA_{1}\bx_{i}$ 
and $\bx_{i,2}=(x_{i(1),2},...,x_{i(d),2})^T=\bA_{2}\bx_{i}$ 
for all $i$,   
$\Var(\bx_{i,s})=\bSig_{s}$ for $s=1,2$. 
Let $\sigma_{(j),s}=\Var(x_{i(j),s})$, $s=1,2$, 
for all $j$. 
Notably, $\sigma_{(j),2}\le \sigma_{(j),1}+\sigma_{(j),2} = \sigma_{(j)}$ for all $j$. 
We assume the following bounded condition, as necessary:
\begin{description}
\item[(C-ii)] 
$\displaystyle \liminf_{d\to \infty} \sigma_{(j),2}>0$
 \ and \   
$\displaystyle \limsup_{d\to \infty} \sigma_{(j)}<\infty$
for all $j$. 
\end{description}
%

The diagonal elements are typically bounded. 
Thus, 
(C-ii) generally holds. 
Under (C-ii), $\sigma_{(j),2}\in (0,\infty)$ as $d\to \infty$ for all $j$. 
Here, for function $f(\cdot)$, ``$f(d) \in (0, \infty)$ as $d\to \infty$'' implies that $\liminf_{d\to \infty}$
$f(d)>0$ and $\limsup_{d\to \infty}f(d)<\infty$. 
%
Then, $\tr(\bSig_{2}^2)/d \ge \sum_{s=1}^d \sigma_{(s),2}^2/d \in (0,\infty)$ and $\tr(\bSig)/d  \in (0,\infty)$
 as $d\to \infty$ under (C-ii). 
Therefore, under (C-i) and (C-ii), 
\begin{equation}
\limsup_{p\to \infty}\lambda_j/d<\infty  \ \mbox{ and } \ 
\liminf_{p\to \infty}\lambda_j/d^{1/2}>0
 \ \mbox{ 
 for $j=1,...,m$}.
\label{SSE2}
\end{equation}
\subsection{Assumptions of high-dimensional distributions}
Let $ \bX-\bmu \bone_n^T =\bH \bLam^{1/2}\bZ$. 
Then, $\bZ$ represents a $d\times n$ sphered data matrix obtained from a distribution with an identity covariance matrix. 
Here, we write $\bZ=(\bz_{1},...,\bz_{d})^T$ and $\bz_{j}=(z_{1j},...,z_{nj})^T,\ j=1,...,d$. 
$E(z_{ij}z_{ij'})=0\ (j\neq j')$ and $\Var(\bz_{j})=\bI_n$. 
For convenience, when $\lambda_{j}=0$ for some $j$, we assume $\Var(\bz_{j})=\bI_n$. 
Let $M_j=\Var(z_{ij}^2)$, $j=1,...,d$. 
If $\bX$ is Gaussian, $z_{ij}$s are i.i.d. as the standard normal distribution $N(0,1)$ and $M_{j}=2$ for all $j$. 
We consider the following assumptions:
\begin{description}
\item[(A-i)] 
$\displaystyle E( z_{ij_1}^2z_{ij_2}^2)=1$,\ $E( z_{ij_1 }z_{ij_2 }z_{ij_3 })=0$,\ 
$E( z_{ij_1} z_{ij_2}z_{ij_3}z_{ij_4})=0$ for all $j_1\neq j_2,j_3,j_4$; \ and 
$\displaystyle \limsup_{d\to \infty} M_{j}<\infty$ for all $j$.
\end{description}

Here, (A-i) naturally holds when $\bX$ is Gaussian. 
\CG{Another example satisfying
(A-i) is the case when $\bX$ has a skew normal distribution (Remark S4.1 in the work by \cite{Aoshima:2018b}).}  
This assumption was provided by \cite{Bai:1996}, \cite{Chen:2010}, and \cite{Aoshima:2011}. 
In contrast, 
\cite{Baik:2006}, \cite{Lee:2010}, and \cite{Yata:2012} 
considered the following assumption:
\begin{equation}
z_{i1},...,z_{id} \ 
\mbox{ are independent (or i.i.d.)}. \label{A'}
\end{equation}
The first assumption in (A-i) is milder than (\ref{A'}). 

Next, we consider a \CG{sub-exponential distribution.} 
Let $\bh_j=({h}_{j(1)},...,{h}_{j(d)})^T$ for all $j$.  
From $x_{i(j),2}=\sum_{s=m+1}^d\lambda_s^{1/2} h_{s(j)}z_{is} $, 
we note that  
$\Var(x_{i(j),2} z_{ij'})=\sigma_{(j),2}$ under (A-i) for $j=1,...,d;\ j'=1,...,m$. 
Under (C-ii), 
we consider the following assumption: 
\CG{
\begin{description}
\item[ (A-ii)] 
There exist positive (fixed) constants $t_1$ and $t_2$ such that
\begin{align*}
&\limsup_{d\to \infty} E\big\{\exp(t x_{i(j),2}^2 ) \big\}<\infty \ \mbox{ for $|t|\le t_1$ and  all $j$; \ and} \\
&\limsup_{d\to \infty} E\big\{\exp(t x_{i(j),2} z_{ij'} ) \big\}<\infty \ \mbox{ for $|t|\le t_2$,  all $j$ and $j'=1,...,m$.}
\end{align*}
\end{description}
}

\CG{
If $\bX$ is Gaussian, $x_{i(j),2}/\sigma_{(j),2}^{1/2}$ follows $N(0,1)$, and $x_{i(j),2}$ and $z_{ij'}$ are independent for all $j$ and $j'=1,...,m$, 
so that 
$
E\{\exp(t x_{i(j),2} z_{ij'} ) \}\le E\{\exp(| t | x_{i(j),2}^2+|t|z_{ij'}^2 ) \}
= E\{\exp(|\sigma_{(j),2} t | x_{i(j),2}^2/\sigma_{(j),2})\}
E\{\exp(| t | z_{ij'}^2)\}=(1-2|\sigma_{(j),2} t | )^{-1/2}(1-2| t | )^{-1/2}
$ 
if $|t|< \min\{1/2, 1/(2\sigma_{(j),2})  \} $ 
for all $j$ and $j'=1,...,m$. 
Thus, when $\bX$ is Gaussian, (A-ii) holds under (C-ii). 
In  Appendix B in the online supplementary material, we consider a milder assumption than (A-ii).}
\subsection{Estimation of eigenvalues and PC directions using the NR method}
Eigenvalue estimation using the NR method is given by (\ref{NR}).
The second term in (\ref{NR}) is the estimator of $\delta$. 
\begin{proposition}[Aoshima and Yata, 2018; Yata and Aoshima, 2013]
\label{pro3.2}
Assume (A-i) and (C-i). Then, it holds for $j=1,...,m$ that $\tilde{\lambda}_{j}/\lambda_{j}=
1+O_P(n^{-1/2})$ and 
\begin{align*}
(n/M_j)^{1/2}  \big({\tilde{\lambda}_{j}}/{\lambda_{j}}-1\big)\Rightarrow N(0,1) \ \mbox{ if \ $\displaystyle \liminf_{d\to \infty}M_j>0$}
\end{align*}
as $d \to \infty$ and $n \to \infty$. 
\end{proposition}
Notably, 
$\tilde{\lambda}_{j}$ is a consistent estimator of $\lambda_{j}$ even when $\delta/\lambda_j \to \infty$. 
Thus, we recommend using $\tilde{\lambda}_{j}$ instead of $\hat{\lambda}_{j}$ for high-dimensional data. 
\cite{Wang:2017} applied $\tilde{\lambda}_{j}$s in ``Shrinkage Principal Orthogonal complEment Thresholding'' 
when $d^{1/2}=o(\lambda_m)$. 
\cite{Anderson:1963} showed that ${n}^{1/2}(\hat{\lambda}_j/\lambda_j-1)\Rightarrow  N(0,2)$ as $n\to\infty$ for Gaussian data when $d$ is fixed. 
Thus, $\tilde{\lambda}_{j}$ has the same limiting distribution as in Proposition \ref{pro3.2} for the Gaussian data. 

When applying the NR method to the PC direction vector, we obtain 
\begin{equation}
\tilde{\bh}_{j}=\{(n-1)\tilde{\lambda}_{j}\}^{-1/2}(\bX-\overline{\bX})\hat{\bu}_{j} 
\ \mbox{ for $j=1,...,n-1$.}
\notag
\end{equation}
Notably, $\tilde{\bh}_{j}=(\hat{\lambda}_{j}/\tilde{\lambda}_{j})^{1/2}\hat{\bh}_{j} $. 
From \cite{Aoshima:2018b}, we obtain the following result. 
\begin{proposition}[Aoshima and Yata, 2018]
\label{pro3.3}
Assume (A-i) and (C-i). 
Then, it holds for $j=1,...,m$ that $\tilde{\bh}_{j}^T\bh_{j}=1+O_P(n^{-1})
$ 
as $d \to \infty$ and $n \to \infty$. 
\end{proposition}

\cite{Aoshima:2018b,Aoshima:2018a} used $\tilde{\bh}_{j}$ to develop a data transformation method under the SSE model in the two-sample test and high-dimensional classification. 
Here, 
$\tilde{\bh}_{j}$ is not a unit vector and $\|\tilde{\bh}_{j}\|^2=\hat{\lambda}_{j}/\tilde{\lambda}_{j}$. 
Further, Angle$(\tilde{\bh}_{j}, \hat{\bh}_{j})=0$, and therefore, Angle$(\tilde{\bh}_{j}, {\bh}_{j})=\mbox{Angle}(\hat{\bh}_{j}, {\bh}_{j})$. 
From Propositions 
\CG{\ref{pro3.3}, E.1 
in Appendix E, and (G.1) in Appendix G} in the online supplementary material, under (A-i) and (C-i), it holds for $j=1,...,m$ that
\begin{align}
&\|\tilde{\bh}_{j}\|^2=1+(\delta/\lambda_j)\{1+O_P(n^{-1/2})\}+O_P(n^{-1}) \ \mbox{ and } \label{NR3} \\
&\|\tilde{\bh}_{j}-\bh_{j}\|^2=(\delta/\lambda_j)\{1+O_P(n^{-1/2})\}+O_P(n^{-1}) 
\notag
\end{align}
as $d \to \infty$ and $n \to \infty$. 
Considering Remark E.1 
in Appendix E by noting that $2\{1-(1+\delta/\lambda_j)^{-1/2}\}<\delta/\lambda_j$, the norm loss of $\tilde{\bh}_{j}$ is larger than that of $\hat{\bh}_{j}$. 
However, from Proposition \ref{pro3.3}, $\tilde{\bh}_{j}$ represents a consistent estimator of $\bh_{j}$ considering the inner product even when $\delta/\lambda_j \to \infty$. 
Further, from Propositions \ref{pro3.3} and (\ref{NR3}), under (A-i) and (C-i), there exists 
a random $d$-dimensional vector $\tilde{ \bv }_j$
such that $ {\bh}_j^T \tilde{ \bv}_j=0$, 
\begin{align}
\tilde{\bh}_j=
\{1+O_P(n^{-1})\}{\bh}_j+\tilde{ \bv}_j \ \mbox{and} \ \| \tilde{ \bv}_j \|^2=(\delta/\lambda_j)\{1+O_P(n^{-1/2})\}+O_P(n^{-1})
\label{NR4}
\end{align}
for $j=1,...,m$. 
The coefficient of ${\bh}_j$ in $\tilde{\bh}_j$ is asymptotically $1$.
In contrast, from (\ref{NR4}), Proposition \ref{pro3.2}, and Preposition E.1,  
it holds that 
\begin{align}
\hat{\bh}_j=(1+\delta/\lambda_j)^{-1/2}\{1+o_P(1)\}\tilde{\bh}_j
=(1+\delta/\lambda_j)^{-1/2}\{1+o_P(1)\}({\bh}_j+ \tilde{ \bv}_j). 
\label{3.2}
\end{align}
The coefficient of ${\bh}_j$ in $\hat{\bh}_j$ depends on noise $\delta$. 
The NR estimator $\tilde{\bh}_j$ has an advantage over $\hat{\bh}_j$ by
applying the two steps described in (\ref{ST1}) and (\ref{ST2}) to 
the threshold estimation of PC directions. 

We consider the following divergence conditions for $d$ and $n$:
\begin{description}
\item[($\star$)] 
$\displaystyle \frac{ \log{d}}{n}=o(1)$ \ as $d\to \infty$ and $n\to \infty$. 
\end{description}
Notably, 
 ($\star$) holds even when $d/n \to \infty$. 
Let $\tilde{\bh}_j=(\tilde{h}_{j(1)},...,\tilde{h}_{j(d)})^T$ 
for all $j$. 
\begin{lemma}
\label{lemma2.1}
Assume (A-i), (A-ii), (C-i), and (C-ii). 
Under ($\star$), it holds for $j=1,...,m$ and $j'=1,...,d$ that 
\begin{align*}
\tilde{h}_{j(j')}={h}_{j(j')}+O_P \Big( ( 
\lambda_{j}^{-1} n^{-1} \log{d})^{1/2} \Big) \ \mbox{ as $d\to \infty$ and $n\to \infty$.
} 
\end{align*}
\end{lemma}

Thus, under the conditions in Lemma \ref{lemma2.1}, we have consistency in the sense that 
$
\tilde{h}_{j(j')}={h}_{j(j')}\{1+o_P(1)\}
$
for $j'$ satisfying 
$
(n \lambda_{j} {h}_{j(j')}^2)^{-1} \log{d}
=o(1).
$
From 
(\ref{3.2}), the elements in $\hat{\bh}_{j}$ are not consistent unless $\delta/\lambda_j=o(1)$.

\section{Automatic sparse PCA methodology}
\setcounter{equation}{0} 
In this section, we propose a thresholded estimator of the PC directions by using the NR method. 
We emphasize that consistency properties of the SPCA methods heavily depend on threshold (tuning) values. 
To overcome the inconvenience, we propose an SPCA method from (\ref{NR3}) and (\ref{NR4}). 

\subsection{Thresholded estimator of PC directions using the NR method}
For the PC direction $\bh_j=({h}_{j(1)},...,{h}_{j(d)})^T$, 
we arrange the elements ${h}_{j(1)},...,{h}_{j(d)}$ in the descending order as 
\begin{align}
|{h}_{oj(1)}|\ge \cdots \ge |{h}_{oj(d)}|. 
\label{h1}
\end{align}
Further, $\sum_{s=1}^d {h}_{oj(s)}^2=1$. 
Under (C-i),  
we assume the following condition for the PC direction as necessary:
\begin{description}
\item[(C-iii)] 
For $j=1,...,m$, there exists an integer $k_{j*}$ (which may depend on $d$), 
 such that 
\begin{align*}
&\sum_{s=1}^{k_{j*}} {h}_{oj(s)}^2 \to 1  \ \mbox{ as $d\to \infty$, \ } \ 
\liminf_{d\to \infty}\lambda_{j} {h}_{oj(k_{j*})}^2>0; \ \mbox{ \ and} \\
&\limsup_{d\to \infty} \frac{|{h}_{oj(k_{j*}+1)}|}{|{h}_{oj(k_{j*})}|}<1 \ \mbox{ when $k_{j*}\le d-1$.}  
\end{align*}
\end{description}
\begin{remark}
\CG{
When $\bSig=\bGamma_d$, $\bh_1$ is $\bone_d/d^{1/2}$, where  $\bGamma_d$ is given by (D.1) in Appendix D in the online supplementary material. 
Then, (C-iii) holds with $k_{1*}=d$ and $m=1$. 
}
\end{remark}

Remark \ref{Rem3} in Section 5.2 and Appendix D 
present some models that satisfy (C-i) and (C-iii). 
From (G.15) in Appendix G in the online supplementary material, under (C-i) through (C-iii), we note that 
$k_{j*} \to \infty$ and  $k_{j*}/\lambda_j \in (0,\infty)$ as $d\to \infty$ for $j=1,...,m$. 
%
Now, we consider removing $\tilde{ \bv}_j$ from $\tilde{\bh}_{j}$ in (\ref{NR4}) for each $j\ (=1,...,m)$. 
Notably, 
$ \|\tilde{ \bv}_j \|\approx \delta/\lambda_j$ from (\ref{NR3}). 
For $\tilde{\bh}_j=(\tilde{h}_{j(1)},...,\tilde{h}_{j(d)})^T$ by the NR method, we arrange the elements $\tilde{h}_{j(1)},...,\tilde{h}_{j(d)}$ in the descending order as 
$\tilde{h}_{oj(1)},..., \tilde{h}_{oj(d)}$, where
\begin{align}
|\tilde{h}_{oj(1)}|\ge \cdots \ge |\tilde{h}_{oj(d)}|.
\notag
\end{align}
Further, $\sum_{s=1}^d \tilde{h}_{oj(s)}^2=\|\tilde{\bh}_{j}\|^2$. 
We consider the following thresholded estimator for an integer $k \in[1,d]$ as 
\begin{align}
\tilde{\bh}_{oj}(k)=(\tilde{h}_{oj(1)},...,\tilde{h}_{oj(k)},0,...,0)^T  
\notag
\end{align}
whose last $d-k$ elements are zero; that is, we replace $\tilde{h}_{oj(k+1)},...,\tilde{h}_{oj(d)}$ in 
$(\tilde{h}_{oj(1)},..., \tilde{h}_{oj(d)})^T$ with $0$. 
Here, we provide the optimal integer $k$ from (\ref{NR3}) and (\ref{NR4}). 
From $\|\tilde{\bh}_{j}\|\ge 1$ w.p.1, 
there exists a unique integer $\tilde{k}_{j} \in [1,d]$ such that 
\begin{align}
\sum_{s=1}^{\tilde{k}_{j}-1}\tilde{h}_{oj(s)}^2<1\ \mbox{ and } \ 
\sum_{s=1}^{\tilde{k}_{j}}\tilde{h}_{oj(s)}^2\ge 1.
\label{ST3}
\end{align}
For a sequence $\{A_s\}$, we define $\sum_{s=1}^0A_s=0$ for convenience, so that $\tilde{k}_{j}=1$ if $\tilde{h}_{oj(1)}^2\ge 1$. 
Notably, 
$\| \tilde{\bh}_{oj}(\tilde{k}_j)\|\approx 1$ and $ \sum_{s=\tilde{k}_{j}+1}^{d}\tilde{h}_{oj(s)}^2\approx \delta/\lambda_j$ for a sufficiently large $d$.  
Thus, we can remove $\tilde{ \bv}_j$ from $\tilde{\bh}_{j}$ in (\ref{NR4}) to obtain the following result. 
We write the elements of $\tilde{\bh}_{oj}(\tilde{k}_{j})$ as 
$\tilde{\bh}_{oj}(\tilde{k}_{j})=(\tilde{h}_{oj*(1)},...,\tilde{h}_{oj*(d)})^T$. 
Further, we adjust the subscript $j'$ of $\tilde{h}_{oj*(j')}$ 
as $\tilde{h}_{j*(j'')}$ corresponding to the order of elements in $\tilde{\bh}_{j}$. 
Let $\tilde{\bh}_{j*}=(\tilde{h}_{j*(1)},....,\tilde{h}_{j*(d)})^T$ for all $j$. 
Let  
\CG{$\eta_{j}=\sum_{s=k_{j*}+1}^{d}{h}_{oj(s)}^2$ 
for $j=1,...,m$.}  
Notably, 
$\eta_{j}\to 0$ as $d\to \infty$ in (C-iii). 
\begin{theorem}
\label{theorem3.1}
Assume (A-i), (A-ii), and (C-i) through (C-iii). 
Under ($\star$), it holds for $j=1,...,m$ that \CG{
$\|\tilde{\bh}_{j*} \|^2=1+O_P(\eta_j+ n^{-1})=1+o_P(1)$ and 
$\tilde{\bh}_{j*}^T{\bh}_{j}=1+O_P(\eta_j+ n^{-1})=1+o_P(1)$} 
as $d \to \infty$ and $n \to \infty$. 
\end{theorem}
 
From Theorem \ref{theorem3.1}, we have the following result. 
\begin{corollary}
\label{corollary3.1}
Assume (A-i), (A-ii), and (C-i) through (C-iii). 
\CG{Under ($\star$), it holds that 
\begin{align*}
&\|\tilde{\bh}_{j*}-\bh_{j} \|^2=
O_P\big(\eta_j+ n^{-1} \big)=o_P(1) \ \mbox{ for $j=1,...,m$;} \\ 
& 
\tilde{\bh}_{j*}^T {\bh}_{j'}=O_P\big(\eta_j^{1/2}+ n^{-1/2}\min\{1,\lambda_j^{1/2}/\lambda_{j'}^{1/2} \} \big)=o_P(1)
 \ \mbox{ and} \\
&\tilde{\bh}_{j*}^T \tilde{\bh}_{j'*}=O_P\big(\eta_j^{1/2}+\eta_{j'}^{1/2}+ n^{-1/2} \big)=o_P(1)
\\
& \mbox{when $m\ge 2$ for $j,j'\le m;\  j\neq  j'$
}
\end{align*}
as $d \to \infty$ and $n \to \infty$. }
\end{corollary}
Thus, $\tilde{\bh}_{j*}$ has the aforementioned consistency properties 
even when $\delta/\lambda_j \to \infty$ without threshold (tuning) values such as $\zeta$ in (\ref{ST1}). 
From Theorem \ref{theorem3.1}, we have 
$
\mbox{Angle}(\tilde{\bh}_{j*},\bh_{j} )=o_P(1)
$
under the conditions in Corollary \ref{corollary3.1}. \CG{
\begin{proposition}
\label{pro4}
Assume (A-i) and (C-i). 
Then, it holds that 
$\tilde{\bh}_{j}^T {\bh}_{j'}=O_P(n^{-1/2}\min\{1,\lambda_j^{1/2}/\lambda_{j'}^{1/2} \} )$ 
as $d \to \infty$ and $n \to \infty$ 
when $m\ge 2$ for $j,j'\le m;\  j\neq j'$. 
\end{proposition}
From Proposition \ref{pro4}, $\tilde{\bh}_{j*}^T {\bh}_{j'}$ and  $\tilde{\bh}_{j}^T {\bh}_{j'}$ 
are of the same order for $j,j'\le m;\  j\neq j'$ if $\eta_j=O( n^{-1/2}\min\{1,\lambda_j^{1/2}/\lambda_{j'}^{1/2} \})$ as $d \to \infty$ and $n \to \infty$. 
}

\begin{remark}
We assume that
$ |\tilde{h}_{oj(\tilde{k}_{j})}|>|\tilde{h}_{oj(\tilde{k}_{j}+1)}|,\ j=1,...,m$ for the sake of simplicity. 
Then, we can simply obtain $\tilde{h}_{j*(j')}$ by
\begin{align}
\tilde{h}_{j*(j')}=
\begin{cases}
\tilde{h}_{j(j')} & \mbox{if $|\tilde{h}_{j(j')}|\ge  |\tilde{h}_{oj(\tilde{k}_{j})}|$
}, \\
0 & \mbox{otherwise} \notag
\end{cases} 
 \quad \mbox{ for $j'=1,...,d$}.
\end{align}
\end{remark}

\subsection{Automatic SPCA}

The computational cost of the SPCA method is 
high because it heavily depends on threshold (tuning) values determined by some cross-validation or information criteria. 
One can automatically yield the threshold estimation with a low computational cost because $\tilde{\bh}_{j*}$ does not depend on any threshold (tuning) values such as $\zeta$ in (\ref{ST1}). 

We call the new PCA method that uses $\tilde{\lambda}_j$s and $\tilde{\bh}_{j*}$s as the ``automatic SPCA (A-SPCA)''. 
\CG{Proposition \ref{pro3.2} provides the details of $\tilde{\lambda}_j$. 
$\tilde{\lambda}_j$s have the consistency as ``$\tilde{\lambda}_{j}/\lambda_j=1+o_P(1)$" even when $n/d\to 0$.}  
The A-SPCA algorithm is given as 
\\[3mm]
{\bf [Automatic SPCA (A-SPCA)]}
\begin{description}
  \item[(Step 1)] Set the dual-sample covariance matrix defined by $\bS_D=(n-1)^{-1}(\bX-\overline{\bX})^T(\bX-\overline{\bX})$. 
  \item[(Step 2)] Find the eigenvalues $\hat{\lambda}_{j}$, and the eigenvectors $\hat{\bu}_j$ of $\bS_{D}$. 
  \item[(Step 3)] Calculate $\tilde{\lambda}_{j}=\hat{\lambda}_{j}-\{\tr(\bS_D)-\sum_{s=1}^{j}\hat{\lambda}_{s}\}/(n-j-1)$ 
  and $\tilde{\bh}_{j}=\{(n-1)\tilde{\lambda}_{j}\}^{-1/2}(\bX-\overline{\bX}) \hat{\bu}_{j}$ for each $j$. 
  Estimate the $j$-th eigenvalue by $\tilde{\lambda}_{j}$.
  \item[(Step 4)] Arrange the elements of $\tilde{\bh}_{j}=(\tilde{h}_{j(1)},...,\tilde{h}_{j(d)})^T$ in the descending order as 
$|\tilde{h}_{oj(1)}|\ge \cdots \ge |\tilde{h}_{oj(d)}|$. 
Find the integer $\tilde{k}_{j}$ such that $\sum_{s=1}^{\tilde{k}_{j}-1}\tilde{h}_{oj(s)}^2<1$ and  
$\sum_{s=1}^{\tilde{k}_{j}}\tilde{h}_{oj(s)}^2\ge 1$.
  \item[(Step 5)] 
Define $\tilde{\bh}_{j*}=(\tilde{h}_{j* (1)},...,\tilde{h}_{j* (d)})^T$ 
by 
\begin{align}
\tilde{h}_{j*(j')}=
\begin{cases}
\tilde{h}_{j(j')} & \mbox{if $|\tilde{h}_{j(j')}|\ge  |\tilde{h}_{oj(\tilde{k}_{j})}|$
}, \\
0 & \mbox{otherwise} \notag
\end{cases} 
 \quad \mbox{ for $j'=1,...,d$}.
\end{align}
  Estimate the $j$-th PC direction using $\tilde{\bh}_{j*}$ for each $j$.   
\end{description}

In Appendix A in the online supplementary material, we describe the application of A-SPCA to estimate $\bSig_1$. 
\CG{ Appendix F presents an R code for A-SPCA.}  
\begin{remark}
\CG{
\cite{Aoshima:2018b} created a data-transformation method based on the strongly spiked eigenstructures 
and proposed a high-dimensional two-sample test using data transformation. 
\cite{Aoshima:2018a,Ishii:2022} proposed high-dimensional classifiers using data transformation. 
The key to these inferences is the estimation of the strongly spiked eigenstructures. 
In future, we apply A-SPCA to these inferences for high-dimensional data. 
}
\end{remark}

\section{Simulation} 
\setcounter{equation}{0} 
In this section, we compare the performance of A-SPCA with the conventional PCA and TSPCA given by (\ref{ST2}). 

First, we set $d=2^{s},\ s=6,...,12$ and $n=\lceil d^{1/2} \rceil$, where $\lceil x \rceil$ denotes the smallest integer $\ge x$.
We generate $\bx_{i}$, $i=1,2,...$ independently from $N_d(\bze, \bSigma)$. 
Then, we consider the following two cases: 
\begin{description}
\item[(S-i)] \ 
$\bSig$ is given by $ \lambda_1=d^{2/3}$, $\lambda_2=d^{1/2}$ and $\lambda_3=\cdots =\lambda_d=1$ 
together with
$\bh_1=(1,0,...,0)^T$ and $\bh_2=(0,1,0,...,0)^T$; 
\item[(S-ii)] \ 
$\bSig$ is given by (D.2) in Appendix D in the online supplementary material  
with $\alpha=0.5$, $\beta=2$, $d_1=\lceil d^{2/3} \rceil$, $d_2=\lceil d^{1/2} \rceil$ and 
$\bOme_{d-d_1-d_2}=\bI_{d-d_1-d_2}$.
\end{description}

 (S-i) satisfies (C-i) and (C-iii) when $m=2$ and $k_{1*}=k_{2*}=1$, and (S-ii) satisfies 
(C-i) and (C-iii) when $m=2$, $k_{1*}=d_1$ and $k_{2*}=d_2$. 
(S-i) does not satisfy (C-ii) because $ \sigma_{(1)}\ge \lambda_1 h_{1(1)}^2\to \infty$. 
We considered (S-i) to handle a considerably sparse PC direction.  

Next, we set $n=10{s},\ s=2,...,10$ and $d=1000$. 
We generated $\bx_{i}$, $i=1,2,...$ independently from the mixture model (\ref{model3}) with $g=2$ and $\varepsilon_1=\varepsilon_2=1/2$ as follows:
\begin{description}
\item[(S-iii)] \ 
For $i=1,2$, $f_i(\bx;\  \bmu_{i},\bPsi_{i})$ is the probability density function of $N_d(\bmu_{i},\bPsi_{i})$, where $\bmu_{1}=(1,...,1,0,...,0)^T$ the first $\lceil d^{2/3} \rceil$ elements are $1$, $\bmu_2=-\bmu_1$, $\bPsi_{1}= ( 0.3^{|i-j|^{1/3}}) $ and $\bPsi_{2}=( 0.4^{|i-j|^{1/3}})$.
\end{description}

Further, 
$\lambda_1\approx d^{2/3}$ and $\bh_1\approx  \bmu_{1}/d^{1/3}$ in (S-iii). 
(S-iii) satisfies (C-i) and (C-iii) when $m=1$. 
Details are provided in Remark \ref{Rem3} in Section 5.2. 
 (S-iii) does not satisfy (A-i). 
Further details can be found in Section 4.1.1 in the work  by \cite{Qiao:2010}. 

Finally, we set $d=2^{s},\ s=6,...,12$ and $n=\lceil d^{1/2} \rceil$.
We consider a considerably non-sparse and non-Gaussian case for the first PC direction as follows: 
\begin{description}
\item[(S-iv)] \ 
We generated $\bx_{i}=\bH\bLam^{1/2}\bz_i$, $i=1,2,...$ independently from $z_{ij}=(y_{ij}-5)/{10}^{1/2}$  $(j=1,...,d)$ 
where $y_{ij}$s are i.i.d. as the chi-squared distribution with $5$ degrees of freedom. 
We set $\bSig=\bGamma_d$ 
where $\alpha=0.5$ and $\beta=1$, and $\bGamma_d$ is given by (D.1) in Appendix D. 
\end{description}

$\lambda_1=0.5d+0.5$ and $\bh_1=d^{-1/2} \bone_d$ in (S-iv). 
Further, (S-iv) satisfies (A-i), (C-i), and (C-iii) when $m=1$ and $k_{1*}=d$. 
However, (S-iv) does not satisfy (A-ii). 
Here, $\bh_1$ appears to be a non-sparse vector in the sense that all elements of $\bh_1$ are nonzero. 

\begin{figure}[h]
\begin{center} 
\includegraphics[width=136mm]{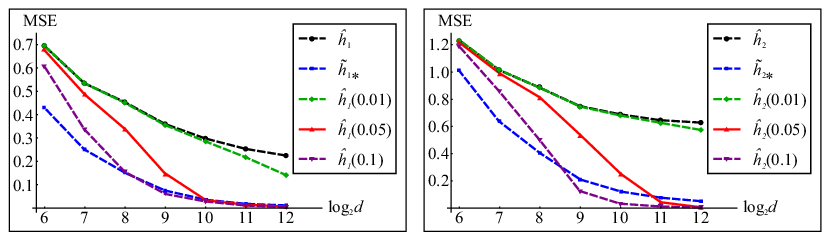}  \\[-2mm]
(S-i) $N_d(\bze, \bSigma)$, $d=2^{s}\ (s=6,...,12)$, where $\bSig$ has $\lambda_1=d^{2/3}$, $\lambda_2=d^{1/2}$, and $\lambda_3=\cdots =\lambda_d=1$ 
together with 
$\bh_1=(1,0,...,0)^T$ and $\bh_2=(0,1,0,...,0)^T$.
\\[2mm]
\includegraphics[width=136mm]{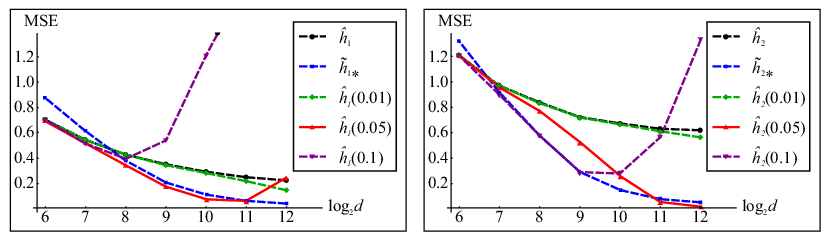}  \\[-2mm]
(S-ii) $N_d(\bze, \bSigma)$, $d=2^{s}\ (s=6,...,12)$, where $\bSig$ has $\lambda_1\approx d^{2/3}$ and $ \lambda_2\approx d^{1/2}$ 
together with 
$\bh_1=(1,...,1,...,0)^T$ whose $\lceil d^{2/3} \rceil$ elements are 1 
and $\bh_2=(0,...,0,1,...,1,0,...,0)^T$, whose $\lceil d^{1/2} \rceil$ elements are 1.
\end{center}
\caption{Average mean-squared errors for PC directions in (S-i) and (S-ii).
 \label{F2}}
\end{figure}

\begin{figure}[h]
\begin{center}
\includegraphics[width=135mm]{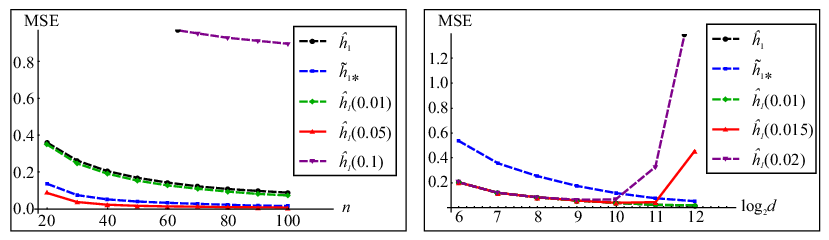}  \\[-3mm]
\hspace{0.2cm}
(S-iii) \hspace{5.6cm}  (S-iv) \hspace{2cm}
 \\[-1mm]
(S-iii) Mixture model (\ref{model3}) with $g=2$, $d=1000$, $n=10{s}\ (s=2,...,10)$, $\lambda_1\approx d^{2/3}$ and $\bh_1\approx  d^{-1/3}(1,...,1,0,...,0)^T$ whose first $\lceil d^{2/3} \rceil$ elements are $d^{-2/3}$.\\
(S-iv) $\bx_i=\bH\bLam^{1/2}\bz_i$; $z_{ij}=(y_{ij}-5)/{10}^{1/2}$  $(j=1,...,d)$ 
in which $y_{ij}$s are i.i.d. as the chi-squared distribution with $5$ degrees of freedom, $d=2^{s}\ (s=6,...,12)$, $\lambda_1=0.5d+0.5$, and $\bh_1=d^{-1/2} \bone_d$. 
\end{center}
\caption{
Average mean-squared errors for the first PC direction in (S-iii) and (S-iv). 
In the left panel, MSE$(\hat{\bh}_{1}(0.1) )$ is too high to describe when $n$ is small.
 \label{F3}}
\end{figure}

We set $\zeta=0.01,\ 0.05$ and $0.1$ for TSPCA using (\ref{ST2}) in (S-i) to (S-iii). 
Further, $\zeta=0.01$ is a soft threshold and $\zeta=0.1$ is a hard threshold.
In (S-iv), we set $\zeta=0.01,\ 0.015$ and $0.02$ for TSPCA because 
all elements of $\bh_1$ are nonzero and small. 
Thus, $\hat{\bh}_{j(\zeta)}$ with a hard threshold results in a considerably poor performance in (S-iv). 
We considered the mean-squared error MSE$(\hat{\bh}_j)=\| \hat{\bh}_j-\bh_j\|^2\ (j=1,...,m)$ and took the average of its outcomes from $2000$ independent replications. 
Similar procedures were performed for 
MSE$(\tilde{\bh}_{j*})$ and MSE$(\hat{\bh}_{j(\zeta)})$. 
We summarized the results in Figs. \ref{F2} and \ref{F3}.

As expected, we observed that A-SPCA achieved preferable performances in (S-i) to (S-iii).  
For (S-iv), the conventional PCA or TSPCA with $\zeta=0.01$ performed better than A-SPCA because all elements of $\bh_1$ are nonzero. 
In addition, $\lambda_1$ is considerably large because $\lambda_1=O(d)$. 
$\hat{\lambda}_1$ is consistent in the sense that $\hat{\lambda}_1/\lambda_1=1+o_P(1)$ (see
Appendix E in the online supplementary material). 
However, A-SPCA performed preferably as $d$ increased, even for the non-sparse case. 
TSPCA with $\zeta=0.015$ and $\zeta=0.02$ exhibited poor performances when 
$d$ is large. 
This is because all elements of $\bh_1=(1/d^{1/2},...,1/d^{1/2})^T$ in (S-iv) become close to zero with an increase in $d$.

Throughout the simulations, TSPCA depended heavily on the threshold value. 
In contrast, A-SPCA performed preferably without using any threshold values.


\section{Shrinkage PC directions and its application to clustering}
\setcounter{equation}{0} 

We consider the shrinkage PC directions using A-SPCA, and we apply them to clustering. 
Fig. \ref{F1} shows that the TSPCA yields a preferable performance even though 
it may not hold the consistency as 
in Corollary \ref{corollary3.1}. 
Further, we show that PCA in the shrinkage PC direction is effective for clustering high-dimensional data.
\CG{We emphasize that the shrinkage PC directions depend on a parameter unlike in A-SPCA.}  

\subsection{Shrinkage PC directions}
For a given constant $\omega_j\in (0,1]$ for each $j\ (=1,...,m)$, we consider the PC shrinkage directions with a cumulative contribution ratio greater than or equal to $\omega_j$. 
For each $j\ (=1,...,m)$, there exists a unique integer $k_{j\omega} \in [1,d] $ 
such that
\begin{equation}
\sum_{s=1}^{k_{j\omega}-1} {h}_{oj(s)}^2< \omega_j \ \mbox{ and } \ 
\sum_{s=1}^{k_{j\omega}} {h}_{oj(s)}^2\ge \omega_j,
\label{6**}
\end{equation}
where ${h}_{oj(s)}$ are given in (\ref{h1}).  
We assume that
$ |{h}_{oj({k}_{j\omega})}|>|{h}_{oj({k}_{j\omega}+1)}|$ 
for simplicity. 
We define 
\begin{align}
{h}_{j\omega (j')}=
\begin{cases}
{h}_{j(j')} & \mbox{if $|{h}_{j(j')}|\ge  |{h}_{oj(k_{j\omega})}|$
}, \\
0 & \mbox{otherwise} \notag
\end{cases} 
 \quad \mbox{ for $j'=1,...,d$}.
\end{align}
Let $\bh_{j\omega}=({h}_{j\omega (1)},...,{h}_{j\omega (d)})^T$ for $j=1,...,m$. 
Further, $\|\bh_{j\omega} \|^2=\sum_{s=1}^{k_{j\omega}} {h}_{oj(s)}^2$ $(\ge \omega_j)$ 
if $|{h}_{oj(k_{j\omega})}|>|{h}_{oj(k_{j\omega}+1)}|$.  
We seek to estimate the shrinkage PC direction $\bh_{j\omega}$.  

As in (\ref{ST3}), 
there exists a unique integer $\tilde{k}_{j\omega} \in [1,d]$ such that 
\begin{align}
\sum_{s=1}^{\tilde{k}_{j\omega}-1}\tilde{h}_{oj(s)}^2<\omega_j \ \mbox{ and } \ 
\sum_{s=1}^{\tilde{k}_{j\omega}}\tilde{h}_{oj(s)}^2\ge \omega_j.
\label{6.2}
\end{align}
We assume that
$ |\tilde{h}_{oj(\tilde{k}_{j\omega})}|>|\tilde{h}_{oj(\tilde{k}_{j\omega}+1)}|$ 
for simplicity. 
We define that 
\begin{align}
\tilde{h}_{j\omega (j')}=
\begin{cases}
\tilde{h}_{j(j')} & \mbox{if $|\tilde{h}_{j(j')}|\ge  |\tilde{h}_{oj(\tilde{k}_{j\omega})}|$
}, \\
0 & \mbox{otherwise} \notag
\end{cases} 
 \quad \mbox{ for $j'=1,...,d$}.
\end{align}
Let $\tilde{\bh}_{j\omega}=(\tilde{h}_{j\omega (1)},...,\tilde{h}_{j\omega (d)})^T$ for $j=1,...,m$. 
Notably, 
$\tilde{\bh}_{j\omega}=\tilde{\bh}_{j*}$ when $\omega_j=1$. 
Under (C-i), 
we assume the following conditions: 
\begin{description}
\item[(C-iii')] 
For some fixed integers $r_j\ (\ge 0)$, 
$\displaystyle \limsup_{d\to \infty}\frac{|{h}_{oj(k_{j\omega}+r_j+1)}|}{|{h}_{oj(k_{j\omega}+r_j)}|}<1
$ 
\ and \ 
$\displaystyle
\liminf_{d\to \infty}\lambda_{j} {h}_{oj(k_{j\omega}+r_j)}^2>0
$, 
\ and 
$\displaystyle \omega_j\lambda_{j}\to \infty $ as $d\to \infty$ for $j=1,...,m$. 
\end{description}
From (G.47) 
in Appendix G in the online supplementary material, under (C-i), (C-ii), and (C-iii'), we note that 
$k_{j\omega}\to \infty$ and $k_{j\omega}/(\omega_{j} \lambda_{j})\in (0,\infty)$ as $d\to \infty$ for $j=1,...,m$. 
\begin{theorem}
\label{thm6}
Assume (A-i), (A-ii), (C-i), (C-ii), and (C-iii'). 
Under ($\star$), it holds for $j=1,...,m$ that \CG{ 
$\|\tilde{\bh}_{j\omega}-{\bh}_{j\omega} \|^2=O_P\{\omega_j(k_{j\omega}^{-1}+n^{-1/2})\}=o_P(\omega_j)$ 
} as $d \to \infty$ and $n \to \infty$. 
\end{theorem}

Thus, $\tilde{\bh}_{j\omega}$ is consistent even when $\delta/\lambda_j\to \infty$. 
However, we cannot construct a consistent estimator of ${\bh}_{j\omega}$ using 
$\hat{\bh}_j$ unless $\delta/\lambda_j=o(1)$. 
The reason is explained in (\ref{3.2}). 

We propose shrinkage PC (SH-PC) scores using the shrinkage PC directions as 
\begin{align}
(\bx_i-\bar{\bx})^T\tilde{\bh}_{j\omega},\ \mbox{ $i=1,...,n$ }
\label{FS3}
\end{align}
for each $j$.
In Section 6, 
we investigated the performance of SH-PC scores for real datasets. 
\begin{remark}
One can use a normalized shrinkage PC direction $\tilde{\bh}_{j\omega}/\|\tilde{\bh}_{j\omega}\|$ in (\ref{FS3}). 
\end{remark}

\subsection{Application to clustering}

For the population distribution of $\bx_i$, 
we consider a $g \ (\ge 2)$-class mixture model whose probability density function is given by
\begin{align}
f(\bx_i)=\sum_{s=1}^g \varepsilon_s f_s(\bx_i;\ \bmu_{s},\bPsi_{s}),\label{model3}
\end{align}
where $\varepsilon_s \in (0,1),\ s=1,...,g,\ \sum_{s=1}^g\varepsilon_s=1$ and 
$f_s(\bx_i;\ \bmu_{s},\bPsi_{s})$ represents the probability density function of a $d$-variate population $\Pi_s$ with mean $\bmu_s$ and covariance matrix $\bPsi_{s}$. 

$E(\bx_i)\ (=\bmu)=\sum_{s=1}^g \varepsilon_s\bmu_s$ and $\Var(\bx_i)=\sum_{s=1}^{g-1}\sum_{s'=s+1}^{g}\varepsilon_s\varepsilon_{s'}(\bmu_{s}-\bmu_{s'})(\bmu_{s}-\bmu_{s'})^T+\sum_{s=1}^g\varepsilon_s \bPsi_{s}\ (=\bSig)$. 
\\

[{\bf Two-class mixture model}]
\\
We assume $g=2$. 
Let $\bmu_{12}=\bmu_1-\bmu_2$ and 
$\Delta_{12}=\|\bmu_{12} \|^2$. 
Then, 
the covariance matrix of the mixture model is given by
$
\bSig=\varepsilon_1\varepsilon_2 \bmu_{12}\bmu_{12}^T+\varepsilon_1\bPsi_1+\varepsilon_2\bPsi_2
$. 
We assume $\bh_1^T\bmu_{12} \ge 0$ without loss of generality. 
If 
\begin{equation}
\lambda_{\max}(\bPsi_s)/\Delta_{12}\to 0 \ \mbox{ as $d\to \infty$ for $s=1,2$}, 
\label{clu2}
\end{equation}
it holds that 
\begin{equation}
\lambda_1/(\varepsilon_1\varepsilon_2\Delta_{12} )\to 1 \ \mbox{ and } \ 
\mbox{Angle}(\bh_1,\bmu_{12})\to 0 \ \mbox{ as $d\to \infty$,} 
\label{clu3}
\end{equation}
where $\lambda_{\max}(\bM)$ denotes the largest eigenvalue of any positive-semidefinite matrix $\bM$. 
Thus, we can obtain a threshold estimation of $\bmu_{12}/\Delta_{12}^{1/2}$ using A-SPCA. 
Furthermore, for the first (true) PC score, it follows that 
\begin{eqnarray}
\bh_1^T(\bx_i-\bmu )= \left\{ \begin{array}{ll}
 \lambda_1^{1/2}\{(\varepsilon_2/\varepsilon_1)^{1/2}+o_P(1)\} & \mbox{ when } \bx_i\in \Pi_1, \\[1mm]
-\lambda_1^{1/2}\{(\varepsilon_1/\varepsilon_2)^{1/2}+o_P(1)\} & \mbox{ when } \bx_i\in \Pi_2 
\end{array} \right. 
\label{clu4}
\end{eqnarray}
as $d\to \infty$ under (\ref{clu2}). 
Hence, one can cluster $\bx_i$s into two groups based on the sign of the estimated first PC scores, as shown in 
Fig. \ref{F1} (i). 
Sections 2 and 3 in the work by \cite{Yata:2019} provide further details on (\ref{clu3}) and (\ref{clu4}). 

Let $\bmu_{12}=(\mu_{(1)},...,\mu_{(d)})^T$. 
We assume $|\mu_{(1)}|\ge \cdots \ge |\mu_{(d)}|$ without loss of generality. 
\begin{remark}
If it holds that
\begin{equation}
\mu_{(k+1)}=\cdots =\mu_{(d)}=0, \ \ \liminf_{d\to \infty}|\mu_{(k)}|>0  \ \mbox{ and } \ k\ge d^{1/2} \ \mbox{for some integer $k$,}
\notag
\end{equation}
(C-i) and (C-iii) with $m=1$ are satisfied when $\max_{s=1,2}\tr(\bPsi_s^2)=O(d)$ and $\bPsi_s$s have NSSE models. 
\label{Rem3}
\end{remark}

For an integer $k$, 
we write $\bx_{i(k)}=(x_{i(1)},...,x_{i(k)})^T$ for all $i$,  
Var$(\bx_{i(k)})=\bSig_{(k)}$, and Var$(\bx_{i(k)})=\bPsi_{s(k)}$ when 
$\bx_{i} \in \Pi_s$. 
Let $\bh_{1(k)}=(h_{1(1)},....,h_{1(k)})^T$, $\bmu_{12(k)}=(\mu_{(1)},...,\mu_{(k)})^T$, and $\Delta_{12(k)}=\| \bmu_{12(k)}\|^2$. 
As in (\ref{clu3})--(\ref{clu4}), 
under the condition that 
\begin{equation}
\lambda_{\max}(\bPsi_{s(k)})/\Delta_{12(k)}\to 0 \ \mbox{ as $d\to \infty$ for $s=1,2$}, 
\label{clu6}
\end{equation}
the following holds: $\mbox{Angle}(\bh_{1(k)},\bmu_{12(k)})\to 0$ as $d\to \infty$. 
Furthermore, 
for the first PC score, it follows that as $d\to \infty$
\begin{eqnarray}
\bh_{1(k)}^T(\bx_i-\bmu )= \left\{ \begin{array}{ll}
\{ \lambda_{\max}(\bSig_{(k)})\}^{1/2} \{(\varepsilon_2/\varepsilon_1)^{1/2}+o_P(1)\} & \mbox{ when } \bx_i\in \Pi_1, \\[1mm]
- \{\lambda_{\max}(\bSig_{(k)})\}^{1/2}\{(\varepsilon_1/\varepsilon_2)^{1/2}+o_P(1)\} & \mbox{ when } \bx_i\in \Pi_2. 
\end{array} \right. 
\notag
\end{eqnarray}
%

Thus, the shrinkage (true) scores were consistent with those under (\ref{clu6}). 
If 
\begin{equation}
\liminf_{d\to \infty}\Delta_{12(k)}/\Delta_{12}>0 \ \mbox{ and } \ \lambda_{\max}(\bPsi_{s(k)})=o\{ \lambda_{\max}(\bPsi_{s})\}, 
\label{clu7}
\end{equation}
(\ref{clu6}) is milder than that in (\ref{clu2}). 
The second condition in (\ref{clu7}) often holds when $k\ll d$. 
Further, from (\ref{clu3}), 
it holds that $\bh_{1(k)}\approx \bmu_{12(k)}/\Delta_{12}^{1/2}$,  
so that 
$$
\frac{\Delta_{12(k)}}{\Delta_{12}}\approx 
\|\bh_{1(k)}\|^2= \sum_{s=1}^k h_{oj(s)}^2.
$$
On the basis of (\ref{6**}) and (\ref{clu7}), if $\liminf_{d\to \infty} \omega_j>0$, 
the SH-PC scores were consistent under milder conditions than those for the PC scores. 
The SH-PC scores by $\tilde{\bh}_{j\omega}$ effectively cluster $\bx_i$s into two groups (see Fig. \ref{F1} (iii) and Section 6.1). 
\\

[{\bf $g\ (\ge 3)$-class mixture model}]
\\
Let $\bmu_{23}=\bmu_2-\bmu_{3}$. 
When $g=3$, 
from Theorem 2 and Lemma 2 in \cite{Yata:2019},  
under certain regularity conditions, 
it holds that 
$\mbox{Angle}(\bh_1,\bmu_{12})\to 0$ and 
$\mbox{Angle}(\bh_2,\bmu_{23})\to 0$ as $d\to \infty$. 
Furthermore, for (true) PC scores, it follows that 
\begin{align}
&
\bh_1^T(\bx_i-\bmu )= \left\{ \begin{array}{ll}
\lambda_1^{1/2} \big[\{(1-\varepsilon_1)/\varepsilon_1\}^{1/2}+o_P(1)\big] & \mbox{ when } \bx_i\in \Pi_1, \\[1mm]
-\lambda_1^{1/2}\big[ \{\varepsilon_1/(1-\varepsilon_1)\}^{1/2}+o_P(1) \big] & \mbox{ otherwise } 
\end{array} \right. 
\label{clu9}
\\
\mbox{and}\notag \\
&
\bh_2^T(\bx_i-\bmu )=
\left\{ \begin{array}{ll}
o_P(\lambda_2^{1/2} ) & \mbox{ when } \bx_i\in \Pi_1, 
\\[1mm]
 \lambda_2^{1/2}\big[ [\varepsilon_3/\{\varepsilon_2(1-\varepsilon_1)\}]^{1/2}+o_P(1)\big] & \mbox{ when } \bx_i \in  \Pi_2, \\[1mm]
-
\lambda_2^{1/2}\big[
[\varepsilon_2/\{\varepsilon_3(1-\varepsilon_1)\}]^{1/2}+o_P(1)\big] & \mbox{ when } \bx_i \in  \Pi_3
\end{array} \right. 
\label{clu10}
\end{align}
for $i=1,...,n$. 
Thus, as in the case of $g=2$, the SH-PC scores by $\tilde{\bh}_{j\omega}$ with $\liminf_{d\to \infty} \omega_j>0$
can effectively cluster $\bx_i$s into three groups (see Section 6.2). 

When $g\ge 4$, \cite{Yata:2019} provided the consistency properties of the PC scores. 

\section{Example: Clustering}

We assess the performance of clustering based on SH-PC scores using (\ref{FS3}). 
We choose $\omega_1=\omega_2 \ (=\omega,\ \mbox{say})$ in (\ref{6.2}) as $\omega=0.1,\ 0.2$ and $0.5$. 
\subsection{Two-class mixture model}
\begin{figure}[h]
\begin{center}
\includegraphics[width=90mm]{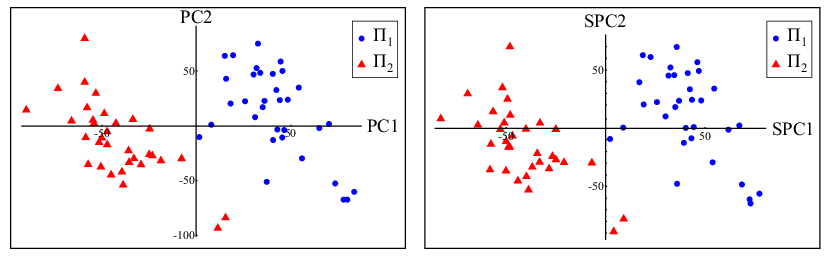}  \\[-3mm]
{\footnotesize \quad \quad
(i) Conventional PCA \hspace{1.3cm} (ii) A-SPCA \hspace{1.4cm} 
 } \\[3mm]
\includegraphics[width=135mm]{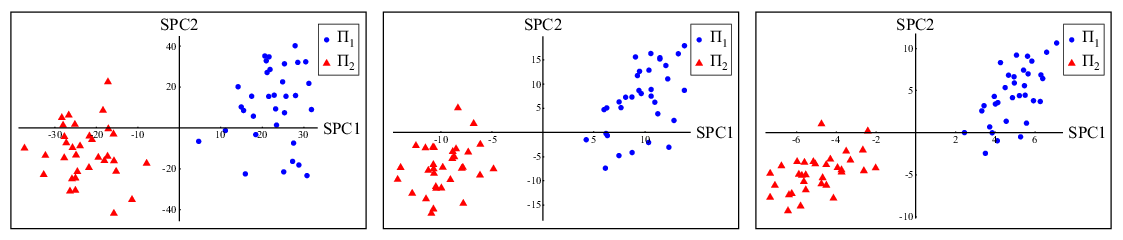}  \\[-3mm]
{\footnotesize 
(iii)  SH-PCs with $\omega=0.5$ \hspace{0.27cm} (iv) SH-PCs with $\omega=0.2$ 
\hspace{0.27cm} (v) SH-PCs with $\omega=0.1$
 }
 \end{center}
\caption{ 
Scatter plots of the first two PC scores for the dataset comprising $\Pi_1$: B-cell with 33 samples and $\Pi_2$: T-cell with 33 samples.  
 \label{F4}}
\end{figure}

We used microarray data provided by \cite{Chiaretti:2004} with $d=12625$ (see Fig \ref{F1}). 
We consider a dataset consisting of
33 samples from $\Pi_1$: B-cell and 
33 samples from $\Pi_2$: T-cell. 
For the mixed $66$ samples, 
 we calculated the first two PC scores using PCA, A-SPCA, and SH-PC with $\omega=0.1,\ 0.2$, and $ 0.5$.
We have that $(\tilde{k}_{1\omega},\tilde{k}_{2\omega})=(17, 50)$, 
$(\tilde{k}_{1\omega},\tilde{k}_{2\omega})=(52, 134)$ and 
$(\tilde{k}_{1\omega},\tilde{k}_{2\omega})=(381, 640)$ for $\omega=0.1,\ 0.2$ and $ 0.5$, respectively.  
Fig. \ref{F4} depicts the scatter plots of the PC scores. 
We observed that the $66$ samples were perfectly classified into two groups based on the sign of the first PC scores 
even when 
the cumulative contribution ratio $\omega$ is as small as $\omega=0.1$. 
These data can be clustered using the SH-PC score with only $17$ significant variables. 
Details are provided in Section 5.2. 

\begin{figure}[h]
\begin{center}
\includegraphics[width=90mm]{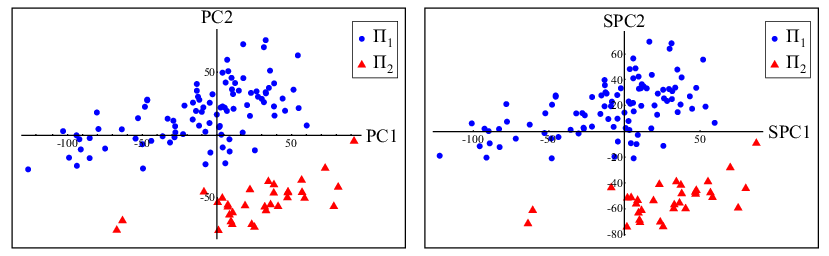}  \\[-3mm]
{\footnotesize \quad \quad
(i) Conventional PCA \hspace{1.3cm} (ii) A-SPCA \hspace{1.4cm} 
 } \\[3mm]
\includegraphics[width=135mm]{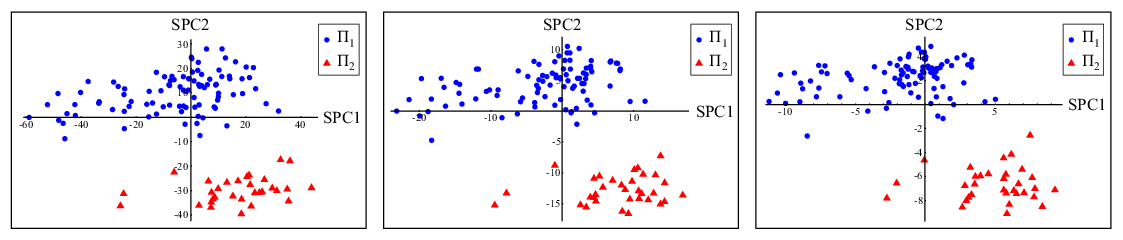}  \\[-3mm]
{\footnotesize 
(iii)  SH-PCs with $\omega=0.5$ \hspace{0.27cm} (iv) SH-PCs with $\omega=0.2$ 
\hspace{0.27cm} (v) SH-PCs with $\omega=0.1$
 }
 \end{center}
\caption{ 
Scatter plots of the first two PC scores for the dataset that comprises $\Pi_1$: B-cell with 95 samples and $\Pi_2$: T-cell with 33 samples.  
 \label{F5}}
\end{figure}

Next, we consider an unbalanced case. 
The B-cells originally contained 95 samples. 
We set $\Pi_1$: B-cell with 95 samples and $\Pi_2$: T-cell with 33 samples. 
As in Fig. \ref{F4}, we present the scatter plots of the PC scores in Fig. \ref{F5}.
The following hold:
 $(\tilde{k}_{1\omega},\tilde{k}_{2\omega})=(88, 13)$, 
$(\tilde{k}_{1\omega},\tilde{k}_{2\omega})=(229, 35)$, and 
$(\tilde{k}_{1\omega},\tilde{k}_{2\omega})=(977, 223)$ for $\omega=0.1,\ 0.2$, and $ 0.5$, respectively.  
The 128 samples were effectively classified into two groups based on the sign of the second SH-PC score; this figure is remarkably different from Fig. \ref{F4}. 
Further,
$\varepsilon_1 \varepsilon_{2}(\bmu_{1}-\bmu_{2})(\bmu_{1}-\bmu_{2})^T$ becomes small in such an imbalanced case, and therefore, $\lambda_{\max}(\bSig_1)$ or $\lambda_{\max}(\bSig_2)$ is probably the largest eigenvalue of $\bSig$ because $\bSigma=\varepsilon_1 \varepsilon_{2}(\bmu_{1}-\bmu_{2})(\bmu_{1}-\bmu_{2})^T+\varepsilon_1\bSigma_1+\varepsilon_2\bSigma_2$.
The second PC direction is probably $(\bmu_1-\bmu_2)/\Delta^{1/2}$. 
Thus, 128 samples were classified into two groups according to the sign of the second SH-PC scores. 
Section 4.3 in the work by \cite{Yata:2019} for further details. 

\subsection{Three-class mixture model}

\begin{figure}[t!]
\begin{center}
\includegraphics[width=90mm]{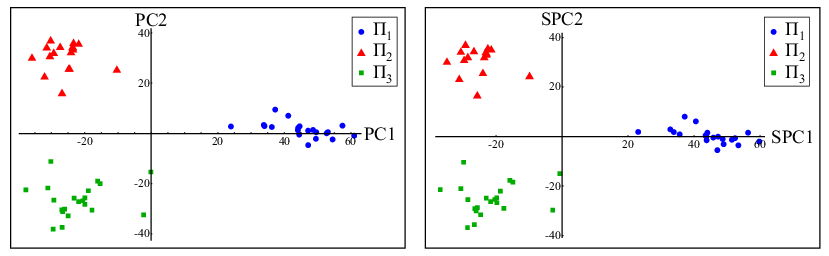}  \\[-3mm]
{\footnotesize \quad \quad
(i) Conventional PCA \hspace{1.3cm} (ii) A-SPCA \hspace{1.4cm} 
 } \\[3mm]
\includegraphics[width=135mm]{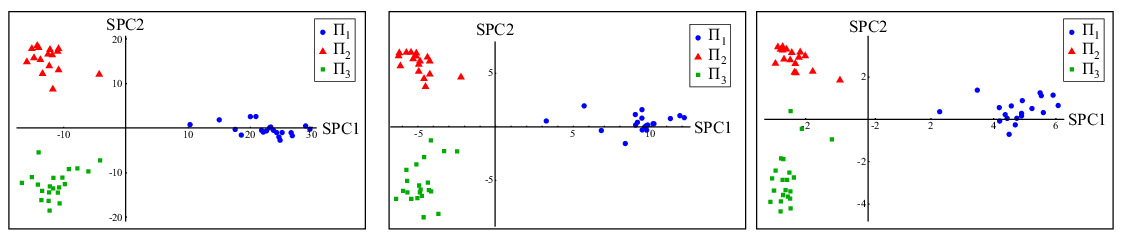}  \\[-3mm]
{\footnotesize 
(iii)  SH-PCs with $\omega=0.5$ \hspace{0.27cm} (iv) SH-PCs with $\omega=0.2$ 
\hspace{0.27cm} (v) SH-PCs with $\omega=0.1$
 }
 \end{center}
\caption{ 
Scatter plots of the first two PC scores for the dataset that comprises $\Pi_1:$ pulmonary carcinoids with 20 samples; 
$\Pi_2:$ normal lung with 17 samples; and
$\Pi_3:$ squamous cell lung carcinomas with 21 samples.  
 \label{F6}}
\end{figure}

We analyzed microarray data from \citet{Bhattacharjee:2001} in which the dataset comprised five lung carcinoma types with $d=3312$. 
We used only three classes: 
$\Pi_1:$ pulmonary carcinoids with 20 samples;
$\Pi_2:$ normal lung with 17 samples; and 
$\Pi_3:$ squamous cell lung carcinoma with 21 samples. 
As in Fig. \ref{F4}, 
we present the scatter plots of the PC scores in Fig. \ref{F6}.
The following hold:
$(\tilde{k}_{1\omega},\tilde{k}_{2\omega})=(17, 16)$, 
$(\tilde{k}_{1\omega},\tilde{k}_{2\omega})=(44, 41)$, and 
$(\tilde{k}_{1\omega},\tilde{k}_{2\omega})=(195, 171)$ for $\omega=0.1,\ 0.2$ and $ 0.5$, respectively. 

We observed that all samples were effectively classified into three groups based on the first two PC scores. 
Further details are provided in (\ref{clu9}) and (\ref{clu10}). 
Fig. \ref{F6} (v) shows that these data can be clustered using the SH-PC score with $(\tilde{k}_{1\omega},\tilde{k}_{2\omega})=(17, 16)$.
Here, the 17 variables in $\bh_{1\omega}$ and the 16 variables in $\bh_{2\omega}$ were all different. 
Therefore, these data can be clustered using the first two SH-PC scores with only $33$ significant variables. 

\section{Conclusion}
\label{sec:conc}
In this study, we investigated TSPCA in high-dimensional settings. 
We proposed a new TSPCA method called automatic SPCA (A-SPCA) and demonstrated that 
it exhibits the consistency property under mild conditions  free from 
any threshold values. 
Therefore, we can quickly obtain a more accurate result at a lower computational cost. 
Further, we proposed shrinkage PC directions and applied them to the clustering. 
We demonstrated the performance of clustering based on the shrinkage PC directions. 
We demonstrated that the datasets could be clustered using a
set of significant variables.


\section*{Supplementary Materials}

We give an estimate of 
the intrinsic part in $\bSig$ using A-SPCA,  
examples of the strongly spiked eigenstructures, 
\CG{asymptotic results for A-SPCA under a milder assumption than (A-ii), 
comparisons between TSPCA and RSPCA,}  
asymptotic properties of the conventional PCA, 
an R code for A-SPCA and proofs of the theoretical results 
in the online supplementary material. 
\par
\section*{Acknowledgements}
\CG{We are very grateful to the associate editor and the reviewers for their constructive comments.} 
The research of the first author was partially supported by Grant-in-Aid for Scientific Research (C), Japan Society for the Promotion of Science (JSPS), under Contract Number 22K03412. 
The research of the second author was partially supported by Grants-in-Aid for Scientific Research (A) and Challenging Research (Exploratory), JSPS, under Contract Numbers 20H00576 and 22K19769. 
\par


\newpage
\vspace{.55cm}
 \centerline{{\LARGE \bf Supplementary Material}}
\vspace{.55cm}

\section*{Appendix A: Estimation of the intrinsic part in the covariance matrix} 
\def\theequation{A.\arabic{equation}}
\def\thesection{A}

In this section, we consider estimating the intrinsic part, $\bSig_{1}=\sum_{s=1}^m \lambda_s \bh_s \bh_s^T$. 
\cite{Fan:2013} 
proposed a covariance matrix estimation procedure called the POET. 
The key point of this procedure is the estimation of $\bSig_{1}$. 
However, they considered an estimation based on $(\hat{\lambda}_j,\hat{\bh}_{j})$s 
such that the estimation does not hold consistency properties unless $\delta/\lambda_m=o(1)$.
See Proposition A.2 in Section A.2.
We apply A-SPCA to 
the estimator of $\bSig_{1}$. 

\subsection{Estimation of scaled PC directions}
Let $\bbeta_j=\lambda_{j}^{1/2}\bh_j$ for $j=1,...,m$. 
Note that 
$\|\bbeta_j\|^2=\lambda_{j}$ for $j=1,...,m$. 
%
Let $\tilde{\bbeta}_{j}=\tilde{\lambda}_j^{1/2}\tilde{\bh}_{j*}$ for $j=1,...,m$. 
By combining Proposition 1 and Theorem 1, 
we have the following result. 
\\
\noindent
{\bf Corollary A.1.} 
{\it 
Assume (A-i), (A-ii), and \CG{ (C-i) to (C-iii). 
Under ($\star$), it holds for $j=1,...,m$ that 
$\|\tilde{\bbeta}_{j}-\bbeta_j \|^2/\lambda_j=O_P(\eta_j+n^{-1})=o_P(1)$} 
as $d\to \infty$ and $n\to \infty$.}  

Let $\hat{\bbeta}_{j}=\hat{\lambda}_j^{1/2}\hat{\bh}_{j}$ for $j=1,...,m$. 
\\
\noindent
{\bf Proposition A.1.} 
{\it 
Assume (A-i) and (C-i). 
Then, it holds for $j=1,...,m$ that 
$\|\hat{\bbeta}_{j}-\bbeta_j \|^2/\lambda_j=\delta/\lambda_j+O_P(n^{-1})$ as $d\to \infty$ and $n\to \infty$.}

Therefore, $\hat{\bbeta}_{j}$ does not hold consistency unless $\delta/\lambda_j=o(1)$. 
In contrast, $\tilde{\bbeta}_{j}$ holds consistency even when $\delta/\lambda_j\to \infty$. 
\subsection{Estimation of $\bSig_{1}$}

The conventional estimator of $\bSig_{1}$ is given by $\widehat{\bSig}_{1}=\sum_{s=1}^m\hat{\bbeta}_{s}\hat{\bbeta}_{s}^T$. 
%
Let $\|\cdot \|_F$ be the Frobenius norm.
\\
\noindent
{\bf Proposition A.2.} 
{\it 
Assume (A-i) and (C-i). Then, it holds that 
$$
\|\widehat{\bSig}_{1}-{\bSig}_{1} \|_F^2=m \delta^2+2\sum_{s=1}^{m}\lambda_s\delta\{1+o_P(1)\} +O_P(\lambda_1^2/n )
 \ \mbox{ as $d\to \infty$ and $n\to \infty$.}
$$
}

Here, $\lambda_{m}$ is the smallest non-zero eigenvalue of $\bSig_1$. 
In addition, $\lambda_1\delta/\lambda_m^2 \ge \delta/\lambda_m$ 
and $\tr(\bSig_2)/d  \in (0,\infty)$ as $d\to \infty$ under (C-ii). 
Therefore, from (2.3), 
we have consistency under the conditions in Proposition A.2 in the sense that 
$\|\widehat{\bSig}_{1}-{\bSig}_{1} \|_F^2=o_P(\lambda_{m}^2)$ if $\lambda_1\delta/\lambda_m^2=o(1)$. 
However, the consistency does not hold unless $\delta/\lambda_m$ approaches $0$. 

Here, we consider the following estimator of $\bSig_{1}$ using A-SPCA:
$$
\widetilde{\bSig}_{1}=\sum_{s=1}^m\tilde{\bbeta}_{s}\tilde{\bbeta}_{s}^T.
$$
See Section S2 in \cite{Aoshima:2018b} for a consistent estimator of $m$. 
\\ 
\noindent
{\bf Theorem A.1.} 
{\it 
Assume (A-i), (A-ii), and (C-i) to (C-iii). 
Under ($\star$), it holds that \CG{
$$
\|\widetilde{\bSig}_{1}-{\bSig}_{1} \|_F^2=
O_P\Big( \lambda_1^2n^{-1}+ \sum_{s=1}^{m}\lambda_{s}^2 \eta_{s} \Big) \ \mbox{ as $d\to \infty$ and $n\to \infty$.}
$$}}

Thus, under the conditions in Theorem A.1, we have consistency in the sense that 
\begin{equation}
\|\widetilde{\bSig}_{1}-{\bSig}_{1} \|_F^2=o_P(\lambda_{m}^2)
 \label{51}
\end{equation}
\CG{if $n^{-1}(\lambda_1^2/\lambda_m^2)=o(1)$ and $\sum_{s=1}^{m}(\lambda_s^2/\lambda_m^2)\eta_{s}=o(1)$.} 
Further, if $\limsup_{d\to \infty} $
$\lambda_1/\lambda_m<\infty$, then (\ref{51}) holds under the conditions in Theorem A.1. 
Therefore, $\widetilde{\bSig}_{1}$ holds the consistency even when $\delta/\lambda_m\to \infty$.

\section*{
\CG{
Appendix B: 
Asymptotic results for A-SPCA under a milder assumption than (A-ii)}
}
\setcounter{equation}{0} 
\def\theequation{B.\arabic{equation}}

\CG{
We consider the following assumption instead of (A-ii):
\begin{description}
\item[ (A-ii')] $\displaystyle \limsup_{d\to \infty}E( x_{i(j),2}^8)<\infty$ \ for all $j$; \ and \ 
$\displaystyle \limsup_{d\to \infty}E\{ (x_{i(j),2} z_{ij'} )^{8} \}<\infty$ \ for all $j$ and $j'=1,...,m$. 
\end{description}
Note that (A-ii) implies (A-ii'). 
Also, if $x_{i(j),2}$s and $z_{ij'}$s are independent, 
the second condition of (A-ii') is met under the first condition of (A-ii') and $\limsup_{d\to \infty}E( z_{ij'}^8)<\infty$ for all $j$. 
We consider the following divergence condition  instead of ($\star$): 
\begin{description}
\item[($\star $')] 
$\displaystyle \frac{ d }{n^4}=o(1)$ \ as $d\to \infty$ and $n\to \infty$. 
\end{description}
Note that ($\star$') holds even when $d/n\to \infty$. 
\\
\noindent
{\bf Proposition B.1.} 
{\it 
After replacing (A-ii) and ($\star$) with (A-ii') and ($\star $'), 
the results in Theorems 1, 2, A.1 and Corollaries 1, A.1 are still justified.
}
}

\setcounter{subsection}{0}
\setcounter{equation}{0}
\CG{
\section*{Appendix C: Comparisons between TSPCA and RSPCA}
}

\def\theequation{C.\arabic{equation}}
\def\thesection{C}

\def\thefigure{C.1}

\CG{
In this section, we present several comparisons between threshold-based SPCA (TSPCA) and regularized SPCA (RSPCA). 
\subsection{Asymptotic property} 
The key of  A-SPCA is the following asymptotic property for the PC-direction by the NR method: 
\begin{align}
\tilde{\bh}_j=\{1+o_P(1)\}\bh_j+\tilde{\bv}_j \ \mbox{  as $d\to \infty$ and $n\to \infty$,}
\label{CC1}
\end{align}
where $\bh_j^T\tilde{\bv}_j=0$; i.e., the coefficient of $\bh_j$ in $\tilde{\bh}_j$ is asymptotically 1. 
See (2.6) for additional details. 
However, to the best of my knowledge, the estimation of PC-directions by RSPCAs  does not hold such a result. 
Based on (\ref{CC1}), we can obtain accurate results for the PC-directions by a thresholded method 
 without requiring threshold values. 
See Section 3.1 for further details. 
Thus the threshold-based estimation has a significant theoretical advantage over RSPCAs. 
\subsection{Computational cost}
\cite{Zou:2006} considered an RSPCA 
under high-dimensional settings. 
The first $g$ PC-directions estimated as follows: 
Let $\bB_1=(\bbeta_1,...,\bbeta_g)$ and $\bB_2$ be $d\times g$ matrices.  
They considered the following optimization problem: 
\begin{align*}
(\widehat{\bB}_1,\widehat{\bB}_2)=&\argmin_{ \bsB_1,\bsB_2 }\sum_{i=1}^n\| \bx_i-\bB_2\bB_1^T \bx_i \|^2+ \psi \sum_{j=1}^g \|\bbeta_j \|^2+  \sum_{j=1}^g \psi_{j} \|\bbeta_j \|_1 \\
&\mbox{subject to $\bB_2^T\bB_2=\bI_g$}, 
\end{align*}
where 
$\psi $ and $\psi_{j}$s are tuning parametors, $\| \bbeta_j\|_1$ is the L1 norm of $\bbeta_j$, and $\widehat{\bB}_1=(\hat{\bbeta}_1,...,\hat{\bbeta}_g)$. 
Then, $\hat{\bbeta}_j/\|\hat{\bbeta}_j\|$ is the estimator of $\bh_j$ for $j=1,...,g$. 
They also expressed the computational cost of the RSPCA as 
$$
 O\{T(dnJ+J^3)\}\ \mbox{ when $d>n$},
$$
where $J$ is the number of nonzero coefficients in the PC-directions and $T$ is the number of iterations before convergence. 
See Section 3.5 in \cite{Zou:2006} for further details. 
On the other hand, by using the singular value decomposition of $\bX-\overline{\bX}$, the computational cost of the TSPCA by (1.5) becomes $O(dn)$. 
The computational cost of the A-SPCA by the R-code in Appendix F is also $O(dn)$. 
We note that because A-SPCA is a TSPCA, it does not require iterations before convergence. 
Overall, TSPCAs are easier to handle than RSPCAs in terms of computational complexity. 
}
 
\CG{
Here, we compared the computational cost of  the A-SPCA with that of the RSPCA  proposed by \cite{Zou:2006} for the (S-i) and (S-ii) settings in Section 4. 
We used the R-code of the RSPCA in the ``elasticnet'', which is available from CRAN (\url{https://cran.r-project.org/web/packages/elasticnet}). 
We set $K=2$ and para=c(0.05,0.05) in the R-code and calculated the ratios of computational costs for the RSPCA over the A-SPCA by 2000 iterations 
for both settings, with the results indicating lower computational costs for the latter, as shown 
in Fig. \ref{FC}. 
\begin{figure}[htbp]
\begin{center} 
\includegraphics[width=136mm]{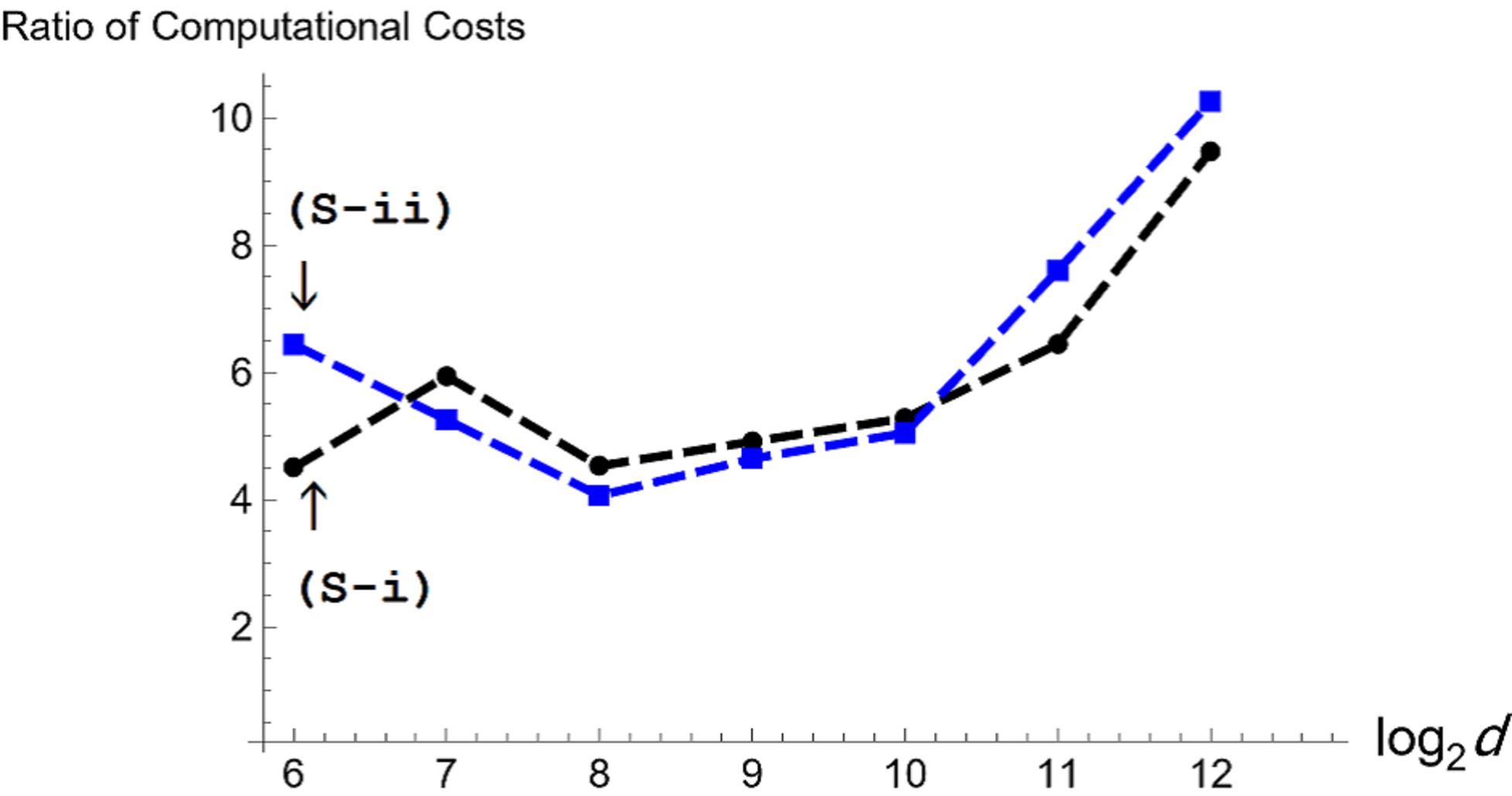} 
\end{center}
\caption{
\CG{
The ratios of the computational costs for the RSPCA over the A-SPCA 
for (S-i) $N_d(\bze, \bSigma)$, $d=2^{s}\ (s=6,...,12)$, $n=\lceil d^{1/2} \rceil$, where $\bSig$ has $\lambda_1=d^{2/3}$, $\lambda_2=d^{1/2}$, and $\lambda_3=\cdots =\lambda_d=1$ 
together with 
$\bh_1=(1,0,...,0)^T$ and $\bh_2=(0,1,0,...,0)^T$, and 
for (S-ii) $N_d(\bze, \bSigma)$, $d=2^{s}\ (s=6,...,12)$, $n=\lceil d^{1/2} \rceil$, where $\bSig$ has $\lambda_1\approx d^{2/3}$ and $ \lambda_2\approx d^{1/2}$ 
together with 
$\bh_1=(1,...,1,...,0)^T$, whose $\lceil d^{2/3} \rceil$ elements are 1 
and $\bh_2=(0,...,0,1,...,1,0,...,0)^T$, whose $\lceil d^{1/2} \rceil$ elements are 1.}
\label{FC}}
\end{figure}
We emphasize that the RSPCA and TSPCA by (1.5) heavily depend on threshold (tuning) values determined 
by specific cross-validation or information criteria. 
In contrast, because the A-SPCA does not depend on any threshold (tuning) values, 
it quickly obtains an accurate result at a lower computational cost. 
}

\section*{Appendix D: Examples of the strongly spiked eigenstructures}
\setcounter{equation}{0} 
\def\theequation{D.\arabic{equation}}

We provide examples of (C-i) and (C-iii). 

First, we consider an intraclass correlation model given by 
\begin{align}
\bGamma_q =\beta(\alpha  \bI_q+(1-\alpha) \bone_{q}\bone_q^T), \label{model1}
\end{align}
where $\alpha \in (0,1)$ and $\beta\ (>0)$ are fixed constants. 
For the model, 
$\lambda_{\max}(\bGamma_q )=\beta\{(1-\alpha)q+\alpha\}$ and the other eigenvalues are $ \alpha \beta$. 
If $\bSig=\bGamma_d$, (C-i) and (C-iii) with $(m, k_{1*})=(1, d)$ are satisfied 
because $\bh_1=d^{-1/2} \bone_d$ when $\bSig=\bGamma_d$.
Then, $\bh_1$ is a non-sparse vector in the sense that all elements of $\bh_1$ are nonzero. 

Next, we consider the following model.
\begin{equation}
\bSig=\left( \begin{array}{ccc}
\bGamma_{d_{1}} &\bO   & \bO \\
\bO & \bGamma_{d_2} &\bO  \\
\bO & \bO & \bOme_{d_{3}}
\end{array} \right)
,
\label{model2}
\end{equation}
where $d_1>d_2>d_3>0$, $d_1+d_2+d_3=d$ and $\bOme_{d_{3}}$ is a 
$d_3$-dimensional non-negative definite matrix. 
If 
$$
d_2 \ge d^{1/2},\ \lambda_{\max} (\bOme_{d_{3}})/d_2 \to 0, \ \mbox{ and } \ 
\{\lambda_{\max} (\bOme_{d_{3}})\}^2/\tr(\bOme_{d_{3}}^2)\to 0 
\ \mbox{ as  $d\to \infty$,} 
$$
(C-i) and (C-iii) with $(m, k_{1*}, k_{2*})=(2, d_1, d_2)$ are met from the fact that 
$\bh_1=(d_1^{-1/2} \bone_{d_1}^T,0,...,0)^T$ and $\bh_2=(0,...,0,d_2^{-1/2} \bone_{d_2}^T,0,...,0)^T$.

\section*{Appendix E: Asymptotic properties of 
the conventional PCA}
\setcounter{equation}{0} 
\def\theequation{E.\arabic{equation}}

Let $v_j=\sum_{i=1}^n(z_{ij}-\bar{z}_{j})^2/(n-1) $ for all $j$, where $\bar{z}_{j}=\sum_{i=1}^nz_{ij}/n$. 
For the conventional PCA, 
we obtain the following results from Proposition 2 and (S6.1) in \cite{Aoshima:2018b}. 
\\
\noindent
{\bf Proposition E.1} (Aoshima and Yata, 2018).  
{\it 
Assume that (A-i) and (C-i). 
It holds for $j=1,...,m$ that 
\begin{align}
&\frac{\hat{\lambda}_{j}}{\lambda_{j}}=v_j+\frac{\delta}{\lambda_j}+O_P(n^{-1})=1+\frac{\delta}{\lambda_j}+O_P(n^{-1/2}) \ \mbox{ and } \notag \\
&\mbox{Angle}(\hat{\bh}_{j}, \bh_{j})=\mbox{Arccos}\bigg(\frac{1}{\sqrt{ 1+\delta/\lambda_j} } +O_P(n^{-1/2})
\bigg) \ \mbox{ as $d\to \infty$ and $n\to \infty$.} \label{3.1}
\end{align}
}
\noindent
{\bf Remark E.1.} 
Equation (\ref{3.1}) is equivalent to 
\begin{align*}
&\hat{\bh}_{j}^T \bh_{j}=(1+\delta/\lambda_j)^{-1/2}+O_P(n^{-1/2}) \ \mbox{ or } \\ 
&\|\hat{\bh}_{j}-\bh_{j}\|^2=2\{1-(1+\delta/\lambda_j)^{-1/2}\}+O_P(n^{-1/2}). \notag 
\end{align*}
 
\citet{Yata:2012} proposed the noise-reduction (NR) methodology to reduce the bias term; it was introduced using a geometric representation of high-dimensional noise. 

\section*{Appendix F: R-code for the A-SPCA}
\setcounter{equation}{0} 
\def\theequation{F.\arabic{equation}}

We give the following R-code for A-SPCA: 
\\
\begin{description}
\item[{\bf Input}]  ASPCA(X, r);  $d\ (\ge 2)$ by $n\ (\ge 4)$ matrix $X$ as $X=(\bx_1,...,\bx_n)$, and  
$r \in [2,\min\{d,n-2\}]$ (the number of components to be computed). 

\item[{\bf Output}]   values$[j]$: The estimator of the $j$-th eigenvalue by A-SPCA (the NR method). \\ 
vectors$[, j]$: The estimator of the $j$-th PC direction by A-SPCA. 
\end{description} 
\begin{lstlisting}[language=R, frame=single]
ASPCA <- function(X,r){
  d <- dim(X)[1]
  n <- dim(X)[2]
  q <- min(n-2, d, r)
  X <- sweep(X, 1, apply(X, 1, mean), '-')
  X0 <- X 
  svd0 <- svd(X0 / (n-1)^(1/2), nu = q, nv = q)
  sval <- svd0$d[1:q]
  svec <- svd0$v[,1:q]
  trSd <- norm(X0, "F")^2 / (n-1)
  nrmval <- numeric(q)
  nrmvec <- matrix(0, d, q)
  aspca <- matrix(0, d, q)
    for (i in 1:q){
      nrmval[i] <- sval[i]^2 - ( trSd 
                    - sum(sval[1:i]^2)) / (n-i-1)
      nrmvec[, i] <- X %*% svec[, i] 
                    / sqrt((n-1) * nrmval[i])
      ord <- order(abs(nrmvec[, i]), decreasing=T)
      cri <- 0
      for (j in 1:d){
        cri <- cri + nrmvec[ord[j], i]^2
        aspca[ord[j], i] <- nrmvec[ord[j], i]
        if (cri >= 1){
          break
      }
    }
  }
  return(list(values=nrmval, vectors=aspca))
}
\end{lstlisting}

\noindent
\CG{
{\bf Remark F.1.} 
One can calculate the shrinkage PC direction $\tilde{\bh}_{j\omega}$ with a given constant $\omega_j\in (0,1]$ 
by using the above code after replacing  ``cri $>=$ 1'' with ``cri $>=$ $\omega_j$''. 
}

\section*{Appendix G: Proofs}
\setcounter{equation}{0} 
\def\theequation{G.\arabic{equation}}

Throughout all the proofs, 
we assume $\bmu=\bze$ for the sake of simplicity. 
Let 
%
$\bu_{j}=(z_{1j},...,z_{nj})^T/(n-1)^{1/2}$ 
and $\dot{\bu}_{j}={\| \bu_{j}\|}^{-1}{\bu_{j}}$ for all $j$. 
From (S6.1) to (S6.5) in Appendix B of \cite{Aoshima:2018b}, 
under (A-i) and (C-i), we have that as $d\to \infty$ and $n\to \infty$
\CG{
\begin{align}
&\tilde{\lambda}_{j}/\lambda_{j}=\|\bu_{j}\|^2+O_P(n^{-1})=1+O_P(n^{-1/2})\notag \\
&\mbox{and} \quad \hat{\bu}_{j}^T\dot{\bu}_{j}=1+O_P(n^{-1})
\ \mbox{ for $j=1,...,m$;}
 \label{D*} \\
&\hat{\bu}_{j'}^T{\bu}_{j}=O_P(n^{-1/2}\min\{1, \lambda_{j'}/\lambda_{j} \} ) 
\ \mbox{ for $j\neq j'\ (\le m)$} \label{D**}.
\end{align}
}
Note that $\|\bu_{j}\|^2=v_j+O_P(n^{-1})$ as $n\to \infty$ for $j=1,...,m$.  
Let $\bP_{n}=\bI_{n}-\bone_{n}\bone_{n}^T/n$. 
Note that $\bone_{n}^T\hat{\bu}_{j}=0$ and $\bP_{n}\hat{\bu}_{j}=\hat{\bu}_{j}$ when $\hat{\lambda}_{j}>0$ since $\bone_{n}^T\bS_{D}\bone_{n}=0$. 
Also, when $\hat{\lambda}_{j}>0$, note that
\begin{align}
(\bX-\overline{\bX})\hat{\bu}_{j}=\bX\bP_n\hat{\bu}_{j}=\bX\hat{\bu}_{j}
=(n-1)^{1/2}\sum_{s=1}^d \lambda_s^{1/2}\bh_s \bu_s^T\hat{\bu}_{j}
.   \label{D} 
\end{align}
Let $z_{ij(j'),x} =z_{ij}x_{i(j'),2}$ for $j=1,...,m;$ $j'=1,...,d$ and all $i$. 
Let 
$\bar{z}_{j(j'),x}=\sum_{i=1}^nz_{ij(j'),x}/n$ and  $\bx_{(j'),2}= (x_{1(j'),2},...,x_{n(j'),2})^T/(n-1)^{1/2}$  for $j=1,...,m;$ $j'=1,...,d$.  
\begin{proof}[Proof of Lemma 1
]
Assume (A-i), (A-ii), 
(C-i), (C-ii) and ($\star$). 
Let $w_{i(j)} =x_{i(j),2}^2-\sigma_{(j),2}$ and $\bar{w}_{(j)}=\sum_{i'=1}^n w_{i'(j)}/n$ for all $i,j$. 
\CG{
From (A-ii) and (C-ii), we note that 
\begin{align*}
\limsup_{d\to \infty}E\big\{\exp(t w_{i(j)})\big\}
=\limsup_{d\to \infty} \frac{ E\big\{\exp(t x_{i(j),2}^2 )\big\}}{ \exp(t \sigma_{(j),2} )}<\infty
\end{align*}
for $|t|\le t_1$ and  all $j$. 
Then, from (A-ii), 
for any $t>0$  satisfying $t=o(1)$ as $d\to \infty$ and $n\to \infty$, 
we have that as $d\to \infty$ and $n\to \infty$
\begin{align}
&P(|\bar{z}_{j(j'),x}| \ge t )\le 2 \exp(-nt^2/ \psi_1)  \ \mbox{ for all $j,j'$; and } \label{D2} \\
&P(|\bar{w}_{(j)}| \ge t )\le 2 \exp(-nt^2/ \psi_2) \ \mbox{ for all $j$.} \label{D3}
\end{align}
for some fixed constants $\psi_1>0$ and  $\psi_2>0$. 
Refer to Section 2.1.3 in 
\cite{Martin:2019} for the details of this results. 
Then, from (\ref{D2}), 
it holds that for $j=1,...,m,$
\begin{align}
\notag
\sum_{j'=1}^d
P\{|\bar{z}_{j(j'),x}|\ge   (2\psi_1 n^{-1}\log{d})^{1/2}  \}&\le 
\sum_{j'=1}^d 2 d^{-2}
\to 0, 
\notag
\end{align}
so that 
\begin{align}
\bar{z}_{j(j'),x}=O_P\{( n^{-1}\log{d})^{1/2}\} \ \mbox{ for all $j,j'$.} 
\label{D4}
\end{align}
}
Similar to (\ref{D4}), from (\ref{D3}), 
we can claim that  
$$
\bar{w}_{(j)}=O_P\{(n^{-1}\log{d})^{1/2}\} 
$$ 
for all $j$, 
so that  
\begin{align}
\sum_{i=1}^nx_{i(j),2}^2/n=\sigma_{(j),2}\{1+o_P(1)\} \ \mbox{ for all $j$}. 
\label{D5}
\end{align}
From (\ref{D4}) and (\ref{D5}),  
we have that 
\begin{align}
 &{\bu}_{j}^T\bx_{(j'),2}=
 O_P\{(  n^{-1}\log{d})^{1/2}\} \  \mbox{ for $j=1,...,m$, and  all $j'$; \ and \ } \notag \\
&\|\bx_{(j),2}\|^2/\sigma_{(j),2}=1+o_P(1) \  \mbox{ for all $j$.}  
\label{D6}
\end{align}
Here, from (\ref{D*}), 
there exists a unit random vector $\bep_{j}$ 
 such that $\dot{\bu}_{j}^T\bep_{j}=0$ and 
\begin{align}
\hat{\bu}_{j}=\{1+O_P(n^{-1})\}\dot{\bu}_{j}+\bep_{j}\times O_P(n^{-1/2}) \ \ 
\mbox{for $j=1,...,m$}.  \label{D7}
\end{align}
Note that $\|\bu_j\|=1+o_P(1)$ as $n\to \infty$ for $j=1,...,m$. 
Then, 
by combining (\ref{D*}), (\ref{D6}) and (\ref{D7}), 
we have for $j=1,...,m$, that 
\begin{equation}
\tilde{\lambda}_{j}^{-1/2}\bx_{(j'),2}^T \hat{\bu}_{j}= O_P\{( {\lambda}_{j}^{-1}  n^{-1}\log{d})^{1/2}\}  
\  \mbox{ for all $j'$.} 
\label{D8}
\end{equation}
On the other hand, 
from (\ref{D*}) and (\ref{D**}), 
we have for $j=1,...,m$, that
\begin{align}
\frac{\bA_1\bX \hat{\bu}_{j}}{\{(n-1)\tilde{\lambda}_{j}\}^{1/2}}&=\tilde{\lambda}_{j}^{-1/2}\sum_{j'=1}^m \lambda_{j'}^{1/2} \|\bu_{j'}\| \bh_{j'} \dot{\bu}_{j'}^T\hat{\bu}_{j} \notag \\
&=
\bh_j\{1+O_P(n^{-1}) \} \notag \\ 
& \quad +O_P\{( n\lambda_j )^{-1/2}\}\times \sum_{j'=1(\neq j)}^m \frac{\lambda_{j'}^{1/2}\bh_{j'}}{ \max\{1, \lambda_{j'}/\lambda_{j}  \}  } 
.
\label{D9}
\end{align}
Note that $\sigma_{(j''),1}\le \sigma_{(j'')}$ and $\sigma_{(j''),1}=\sum_{j'=1}^m\lambda_{j'}h_{j'(j'')}^2$ for all $j''$, 
so that 
\begin{align}
\lambda_{j'}^{1/2}h_{j'(j'')}=O(\sigma_{(j'')}^{1/2})=O(1)
\label{D9*}
\end{align}
for all $j''$ and $j'=1,...,m$. 
Also, note that $\{(n-1)\tilde{\lambda}_{j}\}^{-1/2}\bA_2\bX \hat{\bu}_{j}=\tilde{\lambda}_{j}^{-1/2}(\bx_{(1),2},...,\bx_{(d),2})^T\hat{\bu}_{j}$. 
Then, from (\ref{D}), 
by combining (\ref{D8}) and (\ref{D9}), 
we have for $j=1,...,m$, that
$$
\tilde{h}_{j(j')}={h}_{j(j')}+O_P\{({\lambda}_{j}^{-1}  n^{-1}\log{d})^{1/2}\}  \  \mbox{ for all $j'$.} 
$$
It concludes the result. 
%
\end{proof}

\begin{proof}[Proof of Theorem 1]
Assume (A-i), (A-ii),   
(C-i) to (C-iii) and ($\star$). 
We first consider the proof 
for $\tilde{\bh}_{1*}$. 
We assume $|{h}_{1(1)}|\ge \cdots \ge |{h}_{1(d)}|$ for the sake of simplicity. 
From Lemma 1 
and (C-iii), 
it holds for all $j'$, that 
\begin{align*}
\tilde{h}_{1(j')}^2={h}_{1(j')}^2+o_P\Big({h}_{o1(k_{1*})} \max\{ |{h}_{o1(k_{1*})}|,|{h}_{1(j')}| \}\Big) 
\end{align*}
 as $d\to \infty$ and $n\to \infty$. 
Then, 
we have that 
\begin{align}
&\tilde{h}_{1(j)}^2={h}_{1(j)}^2\{1+o_P(1)\} \ \mbox{ for $j=1,...,k_{1*}$}; \ \mbox{ and}  \notag \\
&\tilde{h}_{1(j)}^2={h}_{1(j)}^2+o_P\big( {h}_{o1(k_{1*})}^2 \big)  \ \mbox{ for $j=k_{1*}+1,...,d$}.
\label{D13}
\end{align}
From $\sigma_{(j),1}=\sum_{j'=1}^m\lambda_{j'}h_{j'(j)}^2$, (C-ii) and (C-iii), 
we note 
that 
\begin{align}
&\lambda_{1} h_{1(j)}^2\in (0,\infty) \ \mbox{ as $d\to \infty$ for $j=1,...,k_{1*}$}; \ \mbox{ and}   \notag \\
&h_{1(j)}^2=O(\lambda_{1}^{-1}) \ \mbox{ for $j=k_{1*}+1,...,d$.}  \label{D13*} 
\end{align}
Also, we note that 
\begin{align}
k_{j'*}\to \infty \ \mbox{ and } \ k_{j'*}/\lambda_{j'}\in (0,\infty)
 \ \mbox{ as $d\to \infty$ for $j'=1,...,m$}.
\label{D13**} 
\end{align}
\CG{
Let $\bh_{1,1}=(h_{1(1)},...,h_{1(k_{1*})},0,...,0)^T$ and $\bh_{1,2}=(0,...,0,h_{1(k_{1*}+1)},...,h_{1(d)})^T$. 
From $\bSig_2^{1/2} \bh_1=\bze$, we note that $\bSig_2^{1/2} \bh_{1,1}=-\bSig_2^{1/2} \bh_{1,2}$, 
so that  
$\bh_{1,1}^T\bSig_2\bh_{1,1}= \bh_{1,2}^T\bSig_2\bh_{1,2}\le  \| \bh_{1,2} \|^2 \lambda_{\max}(\bSig_2)=\eta_{1} \lambda_{m+1}$. 
Here, we have that 
\begin{align*}
 & E\bigg\{ \bigg(  \sum_{s=1}^{k_{1*}}  h_{1(s)} {\bu}_{1}^T\bx_{(s),2} \bigg)^2\bigg\}=O\bigg( \frac{\bh_{1,1}^T\bSig_2\bh_{1,1}}{n}\bigg) =O(\eta_{1} \lambda_{m+1}/n) 
  \  \mbox{ \ and \ } \notag \\
& E \bigg(  \bigg\|\sum_{s=1}^{k_{1*}}  h_{1(s)}\bx_{(s),2}\bigg\|^2 \bigg)=O(\bh_{1,1}^T\bSig_2\bh_{1,1})=O(\eta_{1} \lambda_{m+1}).
\end{align*}
Then, by using Markov's inequality, 
for any $c>0$, 
we have that $P\{ |\sum_{s=1}^{k_{1*}}  h_{1(s)} $ $\times  {\bu}_{1}^T\bx_{(s),2}|^2 \ge c (\eta_{1} \lambda_{m+1}/n) \} \le 
E\{ (  \sum_{s=1}^{k_{1*}}  h_{1(s)} {\bu}_{1}^T\bx_{(s),2} )^2\} n/(\eta_{1} \lambda_{m+1} c) =O(c^{-1}) $ 
and $P( \|\sum_{s=1}^{k_{1*}}  h_{1(s)}\bx_{(s),2}\|^2\ge c \eta_{1} \lambda_{m+1} )=O(c^{-1})$, 
so that 
\begin{align}
\sum_{s=1}^{k_{1*}}  h_{1(s)} {\bu}_{1}^T\bx_{(s),2}=O_P\big( \sqrt{ \eta_{1} \lambda_{m+1}/n}\big) \ \mbox{ and } \ 
\bigg\|\sum_{s=1}^{k_{1*}}  h_{1(s)}\bx_{(s),2} \bigg\|=O_P (\sqrt{ \eta_{1}\lambda_{m+1}}).
\label{D13new} 
\end{align}
By combining
(\ref{D*}), (\ref{D7}) and (\ref{D13new}), we have that 
\begin{align}
 \frac{ \bh_{1,1}^T \bA_2\bX \hat{\bu}_{1}}{\{(n-1)\tilde{\lambda}_{1}\}^{1/2} }
=\tilde{\lambda}_1^{-1/2} \sum_{s=1}^{k_{1*}}  h_{1(s)} \hat{\bu}_{1}^T\bx_{(s),2} =O_P(\eta_{1}^{1/2}n^{-1/2})
=O_P(\eta_{1}+n^{-1}). 
\label{D13new*} 
\end{align}
Note that $|\bh_{1,1}^T\bh_j|=|\bh_{1,2}^T\bh_j|\le \eta_{1}^{1/2} $ for $j\ge 2$. 
Then, from (\ref{D9}), it holds that 
\begin{align}
 \frac{ \bh_{1,1}^T \bA_1\bX \hat{\bu}_{1}}{\{(n-1)\tilde{\lambda}_{1}\}^{1/2} }
&=(1-\eta_{1}) \{1+O_P(n^{-1}) \} +O_P\big( \sqrt{\eta_{1}/n} \big) \notag  \\
&=1+O_P(\eta_{1}+n^{-1} ). 
\label{D13new**}
\end{align}
From (\ref{D13new*}) and (\ref{D13new**}), it holds that 
\begin{align}
\sum_{s=1}^{k_{1*}} h_{1(s)}\tilde{h}_{1(s)}
=1+O_P(\eta_{1}+n^{-1} ). 
\label{D13new***}
\end{align}
We 
note that  
$ E(\sum_{s=1}^{k_{1*}} \bar{z}_{1(s),x}^2)=O(\sum_{s=1}^{k_{1*}}\sigma_{(s),2}/n )=O(k_{1*}/n)$ 
and 
$E( \sum_{s=1}^{k_{1*}}$
$ \|\bx_{(s),2}\|^2  )=O(k_{1*})$. 
Then, from  (\ref{D*}) and (\ref{D13**}), we have that 
$\tilde{\lambda}_1^{-1}\sum_{s=1}^{k_{1*}}$
$ (\bx_{(s),2}^T \bu_1)^2=O_P(1/n)$ 
and 
$\tilde{\lambda}_1^{-1} \sum_{s=1}^{k_{1*}} (\bx_{(s),2}^T  \bep_{1}/n^{-1/2})^2\le \tilde{\lambda}_1^{-1} \sum_{s=1}^{k_{1*}} \|\bx_{(s),2}\|^2/n$
$=O_P(1/n)$, 
where $\bep_{1}$ is defined in (\ref{D7}). 
Thus 
it holds that 
\begin{align}
\tilde{\lambda}_{1}^{-1} \|   (\bx_{(1),2},...,\bx_{(k_{1*}),2})^T\hat{\bu}_{1}\|^2
=O_P(1/n). 
\label{DD1}
\end{align}
Let $\bx_{(j),1}= (x_{1(j),1},...,x_{n(j),1})^T/(n-1)^{1/2}$ for all $j$. 
From (\ref{D9}), (\ref{D9*}) and 
(\ref{D13*}), it holds that 
\begin{align}
\tilde{\lambda}_{1}^{-1} \|   (\bx_{(1),1},...,\bx_{(k_{1*}),1})^T\hat{\bu}_{1}-(h_{1(1)},...,h_{1(k_{1*})} )^T \|^2&=O_P\bigg(\sum_{s=1}^{k_{1*}} (n\lambda_1)^{-1}  \bigg) \notag  \\
&=O_P(1/n). 
\label{DD2}
\end{align}
From (\ref{DD1}) and (\ref{DD2}), it holds that 
$\sum_{s=1}^{{k}_{1*}}(\tilde{h}_{1(s)}-{h}_{1(s)})^2=O_P(1/n)$. 
Then, from (\ref{D13new***}), 
it holds that
\begin{align}
\sum_{s=1}^{{k}_{1*}}\tilde{h}_{1(s)}^2&=
\sum_{s=1}^{{k}_{1*}}\{(\tilde{h}_{1(s)}-{h}_{1(s)})^2+2{h}_{1(s)}\tilde{h}_{1(s)}-{h}_{1(s)}^2\} \notag \\
&=1+O_P(\eta_{1}+ n^{-1}). 
\label{DD3}
\end{align}
Let $\mathcal{D}=\{j|\ \tilde{h}_{1*(j)}=0 \mbox{ for $j=1,...,d$} \}$, 
$\mathcal{D}_1=\{j|\ \tilde{h}_{1*(j)}=0 \mbox{ for $j=1,...,k_{1*}$} \}$, 
$\mathcal{D}_2=\{j|\ \tilde{h}_{1*(j)}=0 \mbox{ for $j=k_{1*}+1,...,d$} \}$ and 
$\mathcal{D}_*=\{k_{1*}+1,...,d \}$. 
From (\ref{D13}) and (C-iii), it holds that 
\begin{align}
\max_{j\in \{k_{1*}+1,...,d\}}\tilde{h}_{1(j)}^2< \min_{j\in \{1,...,k_{1*}\}}\tilde{h}_{1(j)}^2
\label{DD4}
\end{align}
with probability tending to $1$. 
Here, we assume 
$$
\liminf_{d\to \infty,n\to \infty}P(\tilde{k}_1> k_{1*})>0 \ \mbox{ and } \  \liminf_{d\to \infty,n\to \infty}P(\tilde{k}_1\le k_{1*})>0
$$
for the sake of simplicity. 
Then, from (\ref{DD4}), we have that 
\begin{align}
&\mathcal{D}=\mathcal{D}_2  \subset \mathcal{D}_* \ \mbox{ and } \ 
\tilde{h}_{o1(\tilde{k}_1)} \in \{\tilde{h}_{1({k}_{1*}+1) },...,\tilde{h}_{1(d) }   \}
 \ \mbox{ if $\tilde{k}_1> k_{1*}$; \  and} \notag \\
& \mathcal{D}_2= \mathcal{D}_* 
\ 
\mbox{ if $\tilde{k}_1\le k_{1*}$} 
\label{DD5}
\end{align}
with probability tending to $1$. 
%
If $\tilde{k}_1> k_{1*}$, 
from  (2.3), Lemma 1 and  (\ref{DD5}),  
we have that 
\begin{align}
1\le \| \tilde{\bh}_{1*}\|^2 \le 1+\tilde{h}_{o1(\tilde{k}_1)}^2
&=1+O_P(h_{1(k_{1*}+1) }^2+\lambda_1^{-1}n^{-1}\log {d}   ) \notag \\
&=1+O_P(\eta_{1} +n^{-1}). 
\label{DD6}
\end{align}
Note that $\sum_{s \in \mathcal{D}_*\backslash \mathcal{D}_2  } h_{1(s)}^2\le  \sum_{s \in \mathcal{D}_* } h_{1(s)}^2=\eta_{1}$. 
Also, from (\ref{DD5}), note that $\sum_{s \in \mathcal{D}_*\backslash \mathcal{D}_2}\tilde{h}_{1(s)}^2= \| \tilde{\bh}_{1*}\|^2 -\sum_{s=1}^{{k}_{1*}}\tilde{h}_{1(s)}^2$ 
with probability tending to $1$ if $\tilde{k}_1> k_{1*}$. 
Then,  if $\tilde{k}_1> k_{1*}$, from  (\ref{DD3}) and  (\ref{DD6}),  we have that 
\begin{align}
\sum_{s \in \mathcal{D}_*\backslash \mathcal{D}_2  } \tilde{h}_{1(s)}^2=O_P(&\eta_{1} +n^{-1})  \ \mbox{ and}  \notag \\
\bigg|\sum_{s \in \mathcal{D}_*\backslash \mathcal{D}_2  } h_{1(s)}\tilde{h}_{1(s)} \bigg| &\le 
\bigg(\sum_{s \in \mathcal{D}_*\backslash \mathcal{D}_2  } h_{1(s)}^2 \bigg)^{1/2}
\bigg(\sum_{s \in \mathcal{D}_*\backslash \mathcal{D}_2  } \tilde{h}_{1(s)}^2 \bigg)^{1/2}   \notag \\
&=O_P\{\eta_{1}^{1/2}(\eta_{1} +n^{-1})^{1/2}\}
=O_P(\eta_{1} +n^{-1}), \label{DD61}
\end{align}
so that from (\ref{D13new***}), 
\begin{align}
\bh_{1}^T\tilde{\bh}_{1*}=\sum_{s=1}^{k_{1*}} h_{1(s)}\tilde{h}_{1(s)}
+\sum_{s \in \mathcal{D}_*\backslash \mathcal{D}_2  } h_{1(s)}\tilde{h}_{1(s)}
=1+O_P(\eta_{1} +n^{-1}). 
\label{DD7}
\end{align}
When $\tilde{k}_1\le   k_{1*}$, 
from (\ref{DD5}), 
we note that $ \| \tilde{\bh}_{1*}\|^2+\sum_{s \in \mathcal{D}_1} \tilde{h}_{1(s)}^2=\sum_{s=1}^{k_{1*}} \tilde{h}_{1(s)}^2$ 
with probability tending to $1$. 
Then, from $\| \tilde{\bh}_{1*}\|^2\ge 1$, (\ref{D13}) and  (\ref{DD3}) 
if $\tilde{k}_1\le k_{1*}$, we have that 
%
%
\begin{align}
&\sum_{s \in \mathcal{D}_1} \tilde{h}_{1(s)}^2=\sum_{s=1}^{k_{1*}} \tilde{h}_{1(s)}^2- \| \tilde{\bh}_{1*}\|^2
\le \sum_{s=1}^{k_{1*}} \tilde{h}_{1(s)}^2-1 =O_P(\eta_{1} +n^{-1})
  \notag \\ 
&\mbox{and \ } \ 
\sum_{s \in \mathcal{D}_1} |{h}_{1(s)}\tilde{h}_{1(s)}|
=\sum_{s \in \mathcal{D}_1} \tilde{h}_{1(s)}^2\{1+o_P(1)\}=O_P(\eta_{1} +n^{-1}).
\label{DD8*}
\end{align}
Thus from (\ref{D13new***}) and  (\ref{DD3}), 
if $\tilde{k}_1\le k_{1*}$, we have that 
\begin{align}
&\bh_{1}^T\tilde{\bh}_{1*}=\sum_{s=1}^{k_{1*}}{h}_{1(s)}\tilde{h}_{1(s)}-\sum_{s \in \mathcal{D}_1} {h}_{1(s)}\tilde{h}_{1(s)}=1+O_P(\eta_{1} +n^{-1}) \notag \\
&\mbox{ and } \ 
\| \tilde{\bh}_{1*}\|^2=
\sum_{s=1}^{k_{1*}} \tilde{h}_{1(s)}^2-\sum_{s \in \mathcal{D}_1} \tilde{h}_{1(s)}^2=
1+O_P(\eta_{1} +n^{-1}). 
\label{DD9}
\end{align}
If $P(\tilde{k}_1\le k_{1*})=o(1)$ or $P(\tilde{k}_1> k_{1*})=o(1)$, we can obtain (\ref{DD6}) and (\ref{DD7}) or  (\ref{DD9}). 
Thus 
from 
(\ref{DD6}), (\ref{DD7}) and  (\ref{DD9}), 
we can conclude the results for $\tilde{\bh}_{1*}$. 
As for $\tilde{\bh}_{j*}$ with $j\ge 2$, 
we obtain the results similarly. 
It concludes the results of Theorem 1. 
}
\end{proof}

\begin{proof}[Proof of Corollary 1]
\CG{
Assume (A-i) and (A-ii), 
(C-i) to (C-iii) and ($\star$). 
From Theorem 1, 
we can claim the first result of Corollary 1. 
Next, we consider the second result of Corollary 1. 
Note that $|\tilde{\bh}_{j*}^T{\bh}_{j'}|= |(\tilde{\bh}_{j*}-{\bh}_{j})^T{\bh}_{j'} |\le \|\tilde{\bh}_{j*}-{\bh}_{j} \|$ for 
$j<j'\ (\le m)$. 
Thus from the first result of Corollary 1, it holds that as $d\to \infty$ and $n\to \infty$ 
\begin{align}
\tilde{\bh}_{j*}^T{\bh}_{j'}=O_P(\sqrt{ \eta_{j}+n^{-1}  })=O_P(  \eta_{j}^{1/2} +n^{-1/2}) \ \mbox{ for $j<j'\ (\le m)$.} 
\label{DD10} 
\end{align}
Here, we consider the case of
$\tilde{\bh}_{2*}^T{\bh}_{1}$. 
We assume $|{h}_{2(1)}|\ge \cdots \ge |{h}_{2(d)}|$ for the sake of simplicity. 
Let $\bh_{1,2*}=(h_{1(1)},...,h_{1(k_{2*})},0,...,0)^T$. 
Similar to (\ref{D13*}), we note that
\begin{align}
h_{j'' (s)}^2=O(\lambda_{j''}^{-1}) \ \mbox{ for $s=1,...,d$; $j''=1,...,m$}. 
 \label{DD11} 
\end{align}
From (\ref{D13**}) and (\ref{DD11}), it holds that $\| \bh_{1,2*} \|^2\le \sum_{s=1}^{k_{2*}} h_{1(s)}^2 =O(\lambda_2/\lambda_{1} )$, 
so that 
$$
\bh_{1,2*}^T\bSig_2 \bh_{1,2*}\le \|\bh_{1,2*} \|^2\lambda_{\max}(\bSig_2)=O(\lambda_{m+1}\lambda_2/\lambda_{1}).
$$ 
Then, similar to (\ref{D13new})-(\ref{D13new*}), we can claim that 
\begin{align}
 \frac{ \bh_{1,2*}^T \bA_2\bX \hat{\bu}_{2}}{\{(n-1)\tilde{\lambda}_{2}\}^{1/2} }
=\tilde{\lambda}_2^{-1/2} \sum_{s=1}^{k_{2*}}  h_{1(s)} \hat{\bu}_{2}^T\bx_{(s),2} =O_P( \lambda_2^{1/2}/(n\lambda_{1})^{1/2}). 
\label{DD12} 
\end{align}
From (\ref{D13**}) and (\ref{DD11}), 
we note that  $|\bh_{1,2*}^T\bh_{j''}| \le  \sum_{s=1}^{k_{2*}} |h_{1(s)} h_{j''(s)}| =\{\lambda_2/(\lambda_{1} \lambda_{j''})^{1/2}  \} $ 
for $j''(\neq 2) \le m$. 
Also, note that  $|\bh_{1,2*}^T\bh_2|=|(\bh_1-\bh_{1,2*})^T\bh_2 |  \le    \eta_{2}^{1/2}$. 
Then, from (\ref{D9}), we have that 
\begin{align}
 \frac{\bh_{1,2*}^T \bA_1\bX \hat{\bu}_{2}}{\{(n-1)\tilde{\lambda}_{2}\}^{1/2}}=&O_P(\eta_{2}^{1/2}) \notag \\
&+O_P\bigg(( n\lambda_2 )^{-1/2}\sum_{j''=1(\neq 2)}^m \frac{ \lambda_2/\lambda_{1}^{1/2}
}{ \max\{1, \lambda_{j''}/\lambda_{2}  \}  } \bigg)  \notag \\
=& O_P\{ \eta_{2}^{1/2}+ \lambda_2^{1/2}/(n \lambda_{1})^{1/2}\}.
\label{DD13}
\end{align}
From (\ref{DD12}) and (\ref{DD13}), 
it holds that
\begin{align}
\sum_{s=1}^{k_{2*}} h_{1(s)}\tilde{h}_{2(s)}
= O_P\{ \eta_{2}^{1/2}+\lambda_2^{1/2}/(n\lambda_{1})^{1/2}\}.
\label{DD14}
\end{align}
Here, we assume 
$$
\liminf_{d\to \infty,n\to \infty}P(\tilde{k}_2> k_{2*})>0 \ \mbox{ and } \  \liminf_{d\to \infty,n\to \infty}P(\tilde{k}_2\le k_{2*})>0
$$
for the sake of simplicity. 
Let $\mathcal{G}=\{j|\ \tilde{h}_{2*(j)}=0 \mbox{ for $j=1,...,d$} \}$, 
$\mathcal{G}_1=\{j|\ \tilde{h}_{2*(j)}=0 \mbox{ for $j=1,...,k_{2*}$} \}$, 
$\mathcal{G}_2=\{j|\ \tilde{h}_{2*(j)}=0 \mbox{ for $j=k_{2*}+1,...,d$} \}$ and 
$\mathcal{G}_*=\{k_{2*}+1,...,d \}$. 
Then, similar to  (\ref{DD5}), we can claim that 
\begin{align}
\mathcal{G}=\mathcal{G}_2  \subset \mathcal{G}_* 
 \ \mbox{ if $\tilde{k}_2> k_{2*}$; \  and } \ 
 \mathcal{G}_2= \mathcal{G}_* \ 
\mbox{ if $\tilde{k}_2\le k_{2*}$} 
\label{DD14*}
\end{align}
with probability tending to $1$. 
If $\tilde{k}_2\le k_{2*}$, 
similar to  (\ref{DD8*}), we can claim that 
\begin{align}
&\sum_{s \in \mathcal{G}_{1} }  \tilde{h}_{2(s)}^2=O_P(\eta_{2}+n^{-1}) \ \mbox{ and} \notag \\
&\tilde{\bh}_{2*}^T{\bh}_{1}=\sum_{s=1}^{k_{2*}} h_{1(s)}\tilde{h}_{2(s)}-\sum_{s \in \mathcal{G}_{1} } h_{1(s)}\tilde{h}_{2(s)}. 
\label{DD15}
\end{align}
with probability tending to $1$. 
From (\ref{D13**}) and (\ref{DD11}), 
we note that  $\sum_{s \in \mathcal{G}_{1} }  {h}_{1(s)}^2 \le \sum_{s=1}^{k_{2*}}  {h}_{1(s)}^2=O(\lambda_2/\lambda_1) $. 
Then, from (\ref{DD15}), if $\tilde{k}_2\le k_{2*}$, it holds that 
%
%
\begin{align}
\sum_{s \in \mathcal{G}_{1} }  |{h}_{1(s)} \tilde{h}_{2(s)} |
\le \bigg(
 \sum_{s \in \mathcal{G}_{1} }  {h}_{1(s)}^2 
 \sum_{s \in \mathcal{G}_{1} } \tilde{h}_{2(s)}^2  \bigg)^{1/2}
=O_P\{ (\eta_{2}^{1/2} +n^{-1/2}) \lambda_2^{1/2}/\lambda_{1}^{1/2} \}, \notag 
\end{align}
so that 
from (\ref{DD14}) and (\ref{DD15}), 
\begin{align}
\tilde{\bh}_{2*}^T{\bh}_{1}=\sum_{s=1}^{k_{2*}} h_{1(s)}\tilde{h}_{2(s)}-\sum_{s \in \mathcal{G}_{1} } h_{1(s)}\tilde{h}_{2(s)}
=O_P\{ \eta_{2}^{1/2}+\lambda_2^{1/2}/(n\lambda_{1})^{1/2}\}.
\label{DD16}
\end{align}
If $\tilde{k}_2> k_{2*}$, 
similar to  (\ref{DD61}), we can claim that 
\begin{align}
\sum_{s \in \mathcal{G}_{*}\backslash \mathcal{G}_{2}  } \tilde{h}_{2(s)}^2 =O_P(\eta_2+n^{-1}).
\label{DD17}
\end{align}
Then, by noting that $\sum_{s \in \mathcal{G}_{*} \backslash \mathcal{G}_{2}  } h_{1(s)}^2\le 1$, 
it holds that 
\begin{align}
\bigg|\sum_{s \in \mathcal{G}_{*}\backslash \mathcal{G}_{2}  } h_{1(s)}\tilde{h}_{2(s)} \bigg| &\le 
\bigg(\sum_{s \in \mathcal{G}_{*} \backslash \mathcal{G}_{2}  } h_{1(s)}^2 \bigg)^{1/2}
\bigg(\sum_{s \in \mathcal{G}_{*}\backslash \mathcal{G}_{2}  } \tilde{h}_{2(s)}^2 \bigg)^{1/2}   \notag \\
&=O_P\{ (\eta_{2} +n^{-1})^{1/2}\}.  \label{DD18}
\end{align}
Here, similar to  (\ref{DD3}), from (\ref{DD14*}), 
if $\tilde{k}_2> k_{2*}$, 
we can claim that 
\begin{align}
\sum_{s=1}^{k_{2*}}\tilde{h}_{2(s)}^2=\sum_{s=1}^{k_{2*}}\tilde{h}_{o2(s)}^2=1+O_P(\eta_2+n^{-1}).
\label{DD181}
\end{align}
Then, from (2.5), it holds that 
\begin{align}
\sum_{s=k_{2*}+1}^d\tilde{h}_{o2(s)}^2=\| \tilde{\bh}_2 \|^2-\sum_{s=1}^{k_{2*}}\tilde{h}_{o2(s)}^2=(\delta/\lambda_2)\{1+o_P(1) \}+O_P(\eta_2+n^{-1}). 
\notag
\end{align}
Thus from $\liminf_{d\to \infty}\tr(\bSig_2)/d>0 $,  (\ref{D13**}) and (\ref{DD181}), 
if $\tilde{k}_2> k_{2*}$, $\lambda_2=o(d)$ and $\eta_2=o(n^{-1})$, 
we have that 
\begin{align}
&\sum_{s=k_{2*}+1}^{\tilde{k}_2 } \tilde{h}_{o2(s)}^2\ge  (\tilde{k}_2-k_{2*}) \{\tr(\bSig_2)/(d \lambda_2 n) \}\{1+o_P(1) \} \ \mbox{ and} \notag  \\
&\sum_{s=1}^{k_{2*}}\tilde{h}_{o2(s)}^2=1+O_P(n^{-1}).  \label{DD191}
\end{align}
Let $\# (A)$ denote the cardinality of the set $A$. 
Note that 
$$
\#(\mathcal{G}_{*}\backslash \mathcal{G}_{2} )=\tilde{k}_2-k_{2*}
$$
with probability tending to $1$ if $\tilde{k}_2> k_{2*}$. 
From $\sum_{s=1}^{\tilde{k}_2}\tilde{h}_{o2(s)}^2\ge 1$ and (\ref{DD191}), if $\tilde{k}_2> k_{2*}$, $\lambda_2=o(d)$ and $\eta_2=o(n^{-1})$, 
it holds that 
$
 \tilde{k}_2-k_{2*}=O_P(\lambda_2),
$
so that from (\ref{D13**}) and (\ref{DD17}),  
\begin{align}
\bigg|\sum_{s \in \mathcal{G}_{*}\backslash \mathcal{G}_{2}  } h_{1(s)}\tilde{h}_{2(s)} \bigg| &\le 
\bigg(\sum_{s \in \mathcal{G}_{*} \backslash \mathcal{G}_{2}  } h_{1(s)}^2 \bigg)^{1/2}
\bigg(\sum_{s \in \mathcal{G}_{*}\backslash \mathcal{G}_{2}  } \tilde{h}_{2(s)}^2 \bigg)^{1/2}   \notag \\
&=O_P\{(\lambda_2/\lambda_1)^{1/2} (\eta_{2} +n^{-1})^{1/2}\}.  \label{DD20}
\end{align}
From (2.3), 
note that $\lambda_1/\lambda_2=O(1)$ if $\liminf_{d\to \infty} \lambda_2/d>0$. 
Thus it holds that 
\begin{align}
&\eta_{2}^{1/2}+\lambda_2^{1/2}/(n\lambda_{1})^{1/2}=O(\eta_{2}^{1/2}) \ \mbox{  if  $\displaystyle \liminf_{d\to \infty,n\to \infty} n\eta_2>0$; and} \notag \\
&\eta_{2}^{1/2}+\lambda_2^{1/2}/(n\lambda_{1})^{1/2}=O(\eta_{2}^{1/2}+n^{-1/2}) \ \mbox{  if $\displaystyle \liminf_{d\to \infty} \lambda_2/d>0$.} \label{DD21}
\end{align}
Thus from (\ref{DD14}), (\ref{DD18}), (\ref{DD20}) and (\ref{DD21}), if $\tilde{k}_2> k_{2*}$, 
we have that 
\begin{align}
\tilde{\bh}_{2*}^T{\bh}_{1}=\sum_{s=1}^{k_{2*}} h_{1(s)}\tilde{h}_{2(s)}+\sum_{s \in \mathcal{G}_{*}\backslash \mathcal{G}_{2}  } h_{1(s)}\tilde{h}_{2(s)} 
=O_P\{ \eta_{2}^{1/2}+\lambda_2^{1/2}/(n\lambda_{1})^{1/2}\}.
\label{DD22}
\end{align}
If $P(\tilde{k}_2> k_{2*})=o(1)$ or $P(\tilde{k}_2\le k_{2*})=o(1)$, we can obtain (\ref{DD16}) or (\ref{DD22}). 
Thus from  (\ref{DD16}) and (\ref{DD22}), 
we can conclude the second result for $\tilde{\bh}_{2*}^T{\bh}_{1}$. 
As for $\tilde{\bh}_{j*}^T{\bh}_{j'}$ with $j'<j;\ (j',j)\neq (1,2)$, 
we obtain the result similarly. 
From (\ref{DD10}),  
we can claim the second result of Corollary 1. }

For the third result of Corollary 1, 
by noting that.
\begin{align}
\tilde{\bh}_{j*}^T \tilde{\bh}_{j'*}=&{\bh}_{j}^T \tilde{\bh}_{j'*}+(\tilde{\bh}_{j*}-{\bh}_{j})^T\tilde{\bh}_{j'*} \notag \\
=&
{\bh}_{j}^T \tilde{\bh}_{j'*}+
{\bh}_{j'}^T\tilde{\bh}_{j*}
+
(\tilde{\bh}_{j*}-{\bh}_{j})^T(\tilde{\bh}_{j'*}- {\bh}_{j'})
\notag
\end{align}
for $j\neq j'$, from the second result of Corollary 1, it concludes the result. 
\end{proof}

\CG{
\begin{proof}[Proof of Proposition 3
]
From (\ref{D}), it holds that 
$
\tilde{\bh}_{j}^T\bh_{j'}= \lambda_{j'}^{1/2} \bu_{j'}^T\hat{\bu}_{j}/\tilde{\lambda}_j^{1/2}
$ for $j\neq j'$. 
Thus from (\ref{D*}) and (\ref{D**}), we can conclude the result.
\end{proof}
}
\begin{proof}[Proof of Theorem 2
]
Assume (A-i), (A-ii),  
(C-i), (C-ii), (C-iii') and ($\star$). 
We first consider the proof for $\tilde{\bh}_{1\omega}$. 
We assume $|{h}_{1(1)}|\ge \cdots \ge |{h}_{1(d)}|$ for the sake of simplicity. 
Similar to 
(\ref{D13}), 
we can claim that 
\begin{align}
&\tilde{h}_{1(j)}^2={h}_{1(j)}^2\{1+o_P(1)\} \ \mbox{ for $j=1,...,k_{1\omega}+r_1$}; \ \mbox{ and} \notag
\\
&\tilde{h}_{1(j)}^2={h}_{1(j)}^2+o_P\big( {h}_{o1(k_{1\omega}+r_1)}^2 \big)  \ \mbox{ for $j=k_{1\omega}+r_1+1,...,d$}
\label{D33}
\end{align}
 as $d\to \infty$ and $n\to \infty$. 
Similar to (\ref{D13*}), 
from (C-iii'), we note for $j'=1,...,m$, that
\begin{align}
&\lambda_{1} h_{1(j)}^2\in (0,\infty) \ \mbox{ as $d\to \infty$ for $j=1,...,k_{1\omega}+r_1$}; \notag \\
&\mbox{ and } \ h_{1(j)}^2=O(\lambda_{1}^{-1}) \ \mbox{ for $j=k_{1\omega}+r_1+1,...,d$.}  \label{D331}
\end{align}
Also, we note that 
\begin{align}
k_{j'\omega}\to \infty \ \mbox{ and } \ k_{j'\omega}/(\omega_{j'} \lambda_{j'})\in (0,\infty)
 \ \mbox{ as $d\to \infty$ for $j'=1,...,m$}.
\label{D34} 
\end{align}
Then, 
from (\ref{D33}), 
we have that
\begin{align}
&\omega_1 \le \| {\bh}_{1\omega }\|^2 \le \omega_1+{h}_{o1(k_{1\omega})}^2=\omega_1+O(\lambda_{1}^{-1})=\omega_1
+O(\omega_1/k_{1\omega}) \ \mbox{ and} \notag \\
&\omega_1 \le \| \tilde{\bh}_{1\omega }\|^2 \le \omega_1+\tilde{h}_{o1( \tilde{k}_{1\omega})}^2=
\omega_1
+O_P(\omega_1/k_{1\omega}).
\label{D35}
\end{align}
\CG{
From (C-ii), all the elements of $\bSig_2$ are bounded. 
Thus, from (\ref{D331}) and  ${\bh}_{1\omega }=({h}_{1(1)},...,{h}_{1(k_{1\omega})},0,...,0)^T$,
we note that ${\bh}_{1\omega}^T \bSig_2 {\bh}_{1\omega}=O(k_{1\omega}^2/\lambda_1) $. 
Then, similar to (\ref{D13new})-(\ref{D13new*}), 
from (\ref{D34}), we can claim that 
\begin{align}
\frac{{\bh}_{1\omega }^T \bA_2\bX \hat{\bu}_{1}}{\{(n-1)\tilde{\lambda}_{1}\}^{1/2} }=O_P(k_{1\omega} \lambda_1^{-1} n^{-1/2}) 
=O_P(\omega_1 n^{-1/2}). 
\label{D351} 
\end{align}
From  (\ref{DD11}), 
we note that $|{\bh}_{1\omega }^T\bh_j|=|\sum_{s=1}^{k_{1\omega} }   h_{1(s)} h_{j(s)} |=O\{k_{1\omega}/(\lambda_1\lambda_j)^{1/2} \} $ 
for $j=2,...,m$. 
Then, from (\ref{D9}) and (\ref{D35}), 
it holds that 
\begin{align}
\frac{{\bh}_{1\omega }^T \bA_1\bX \hat{\bu}_{1}}{\{(n-1)\tilde{\lambda}_{1}\}^{1/2} }=
&{\bh}_{1\omega }^T\bh_1\{1+O_P(n^{-1}) \}
+O_P(k_{1\omega} \lambda_1^{-1} n^{-1/2})  \notag \\
=&\omega_1+O_P(\omega_1/k_{1\omega}+ \omega_1 n^{-1/2}). 
\label{D352} 
\end{align}
From (\ref{D351}) and  (\ref{D352}), 
it holds that 
\begin{align} 
\sum_{s=1}^{{k}_{1\omega}}\tilde{h}_{1(s)}{h}_{1(s)}
=\omega_1+O_P(\omega_1/k_{1\omega} 
+ \omega_1n^{-1/2} ).
\label{D353} 
\end{align}
From (\ref{D34}), note that 
$ E(\sum_{s=1}^{k_{1\omega}}  \bar{z}_{1(s),x}^2)=O(\omega_1 \lambda_1/n)$ 
and $E( \sum_{s=1}^{k_{1\omega}} \|\bx_{(s),2}\|^2  )=O(\omega_1 \lambda_1)$. 
Then,  similar to (\ref{DD1})-(\ref{DD2}), we have that 
$\sum_{s=1}^{{k}_{1\omega}}\{(\tilde{h}_{1(s)}-{h}_{1(s)})^2=O_P(\omega_1/n)$, 
so that from (\ref{D33})-(\ref{D35}) and  (\ref{D353}),
\begin{align}
\ \sum_{s=1}^{{k}_{1\omega}+r_1 }\tilde{h}_{1(s)}^2&=
\sum_{s=1}^{{k}_{1\omega}}\{(\tilde{h}_{1(s)}-{h}_{1(s)})^2+2{h}_{1(s)}\tilde{h}_{1(s)}-{h}_{1(s)}^2\}
+\sum_{s={k}_{1\omega}+1}^{{k}_{1\omega}+r_1 }\tilde{h}_{1(s)}^2
 \notag  \\
&=\omega_1+O_P(\omega_1/k_{1\omega}+ \omega_1 n^{-1/2}). 
\label{D36}
\end{align}
From (\ref{D33}), it holds that 
\begin{align}
\max_{j\in \{k_{1\omega}+r_1+1,...,d\}}\tilde{h}_{1(j)}^2< \min_{j\in \{1,...,k_{1*}+r_1\}}\tilde{h}_{1(j)}^2
\label{D361}
\end{align}
with probability tending to $1$. 
Here, we assume 
$$
\liminf_{d\to \infty,n\to \infty}P(\tilde{k}_{1\omega}\ge k_{1\omega}+r_1)>0 \ \mbox{ and } \  \liminf_{d\to \infty,n\to \infty}P(\tilde{k}_{1\omega}< k_{1\omega}+r_1)>0
$$
for the sake of simplicity. 
Then, if $\tilde{k}_{1\omega}\ge k_{1\omega}+r_1$, from (\ref{D353}) and (\ref{D361}), it holds that 
\begin{align}
\tilde{\bh}_{1\omega}^T\bh_{1\omega}=\sum_{s=1}^{{k}_{1\omega}}\tilde{h}_{1(s)}{h}_{1(s)}=
\omega_1+O_P(\omega_1/k_{1\omega} 
+ \omega_1n^{-1/2} ).
\label{D37}
\end{align}
Let $\mathcal{D}_{\omega}=\{j|\ \tilde{h}_{1\omega(j)}=0 \mbox{ for $j=1,...,k_{1\omega}+r_1$} \}$ and 
$\mathcal{G}_{\omega}=\{j|\ \tilde{h}_{1\omega(j)}=0 \mbox{ for $j=1,...,k_{1\omega}$} \}$. 
When $\tilde{k}_{1\omega}< k_{1\omega}+r_1$, from (\ref{D361}), we note that 
$ \| \tilde{\bh}_{1\omega}\|^2+\sum_{s \in \mathcal{D}_\omega} \tilde{h}_{1(s)}^2=\sum_{s=1}^{k_{1\omega}+r_1} \tilde{h}_{1(s)}^2$ 
with probability tending to $1$. 
Then, if $\tilde{k}_{1\omega}< k_{1\omega}+r_1$, from (\ref{D33}), (\ref{D35}) and (\ref{D36}), we have that 
$\sum_{s \in \mathcal{G}_\omega} \tilde{h}_{1(s)}^2\le \sum_{s \in \mathcal{D}_\omega} \tilde{h}_{1(s)}^2=\sum_{s=1}^{k_{1\omega}+r_1} \tilde{h}_{1(s)}^2-\| \tilde{\bh}_{1\omega}\|^2
=O_P(\omega_1/k_{1\omega}+ \omega_1  n^{-1/2} )$ and
$\sum_{j \in \mathcal{G}_\omega}$
$ \tilde{h}_{1(j)} {h}_{1(j)}=\sum_{j \in \mathcal{G}_\omega} \tilde{h}_{1(j)}^2\{1+o_P(1)\}$. 
Thus, if $\tilde{k}_{1\omega}< k_{1\omega}+r_1$, from (\ref{D353}), it holds that 
\begin{align}
\tilde{\bh}_{1\omega}^T\bh_{1\omega}&=\sum_{s=1}^{{k}_{1\omega}}\tilde{h}_{1(s)}{h}_{1(s)}
-\sum_{s \in \mathcal{G}_\omega} \tilde{h}_{1(s)}{h}_{1(s)} \notag \\
&=\omega_1+O_P(\omega_1/k_{1\omega}+ \omega_1n^{-1/2}). 
\label{D38}
\end{align} 
If $P(\tilde{k}_{1\omega}< k_{1\omega}+r_1)=o(1)$ or $P(\tilde{k}_{1\omega}\ge k_{1\omega}+r_1)=o(1)$, 
we can obtain  (\ref{D37}) or (\ref{D38}). 
Thus 
from 
(\ref{D35}), (\ref{D37}) and (\ref{D38}), 
we can conclude the result for $\tilde{\bh}_{1\omega}$.} 
As for $\tilde{\bh}_{j\omega}$ with $j\ge 2$, 
we obtain the results similarly. 
It concludes the result of Theorem 2. 
\end{proof}

\begin{proof}[Proofs of Corollary A.1 
and Proposition A.1
]
For Corollary A.1, 
from Theorem 1 
and (\ref{D*}), 
under the conditions in Corollary A.1, 
we have that for $j=1,...,m,$ that 
$\|\tilde{\bbeta}_j-\bbeta_j\|^2/\lambda_j=v_j\| \tilde{\bh}_{j*} \|^2+1-2v_j^{1/2}\tilde{\bh}_{j*}^T{\bh}_{j}+O_P(n^{-1})$ 
as $d\to \infty$ and $n\to \infty$. 
Then, by noting that $v_j^{1/2}=1/2+v_j/2+O_P(n^{-1})$, from Theorem 1, 
we can conclude the result in Corollary A.1. 
%

For Proposition A.1,  
from Proposition 2 
and (\ref{D*}), under (A-i) and (C-i), 
it holds for $j=1,...,m,$ that 
\begin{align}
\hat{\lambda}_j^{1/2}\hat{\bh}_{j}^T\bh_j=\tilde{\lambda}_j^{1/2}\tilde{\bh}_{j}^T\bh_j&= \lambda_{j}^{1/2}\{v_j^{1/2}+O_P(n^{-1})\}  \notag \\
&=\lambda_{j}^{1/2}\{1/2+v_j/2+O_P(n^{-1})\}.
\label{D21}
\end{align}
Then, from Proposition E.1, 
we can conclude the result in Proposition A.1. 
\end{proof}

\begin{proof}[Proofs of Theorem A.1 
and Proposition A.2
]
We first consider the proof of Theorem A.1. 
From Theorem 1,
(\ref{D*}) and (\ref{D13**}), 
under the conditions in Theorem A.1,  
we have that for $j=1,...,m,$ \CG{that 
\begin{align}
\|\tilde{\bbeta}_j\tilde{\bbeta}_j^T-\bbeta_j\bbeta_j^T\|_F^2
=&\lambda_j^2\{1+2(v_j-1)+O_P(n^{-1}+\eta_j)\}+\lambda_j^2 \notag  \\
&-2\lambda_j^2\{1+(v_j-1)+O_P(n^{-1}+\eta_j) \} \notag \\
=&O_P\{\lambda_j^2( n^{-1}+\eta_j)\} \label{D22}
\end{align}
as $d\to \infty$ and $n\to \infty$.  
Note that 
$|\tr\{(\tilde{\bbeta}_j\tilde{\bbeta}_j^T-\bbeta_j\bbeta_j^T)(\tilde{\bbeta}_{j'}\tilde{\bbeta}_{j'}^T-\bbeta_{j'}\bbeta_{j'}^T)\}|\le 
\|\tilde{\bbeta}_j\tilde{\bbeta}_j^T-\bbeta_j\bbeta_j^T \|_F\cdot \|\tilde{\bbeta}_{j'}\tilde{\bbeta}_{j'}^T-\bbeta_{j'}\bbeta_{j'}^T\|_F\le 
\|\tilde{\bbeta}_j\tilde{\bbeta}_j^T-\bbeta_j\bbeta_j^T \|_F^2+\|\tilde{\bbeta}_{j'}\tilde{\bbeta}_{j'}^T-\bbeta_{j'}\bbeta_{j'}^T\|_F^2
$ for $j\neq j'$. 
Then, it holds that 
$$
\|\widetilde{\bSig}_{1}-{\bSig}_{1} \|_F^2 \le m \sum_{s=1}^m\|\tilde{\bbeta}_s\tilde{\bbeta}_s^T-\bbeta_s\bbeta_s^T\|_F^2. 
$$
Thus from (\ref{D22}), we can conclude the result in Theorem A.1.}  

For Proposition A.2, 
from Proposition E.1 and (\ref{D21}), 
under (A-i) and (C-i), 
we have for $j=1,...,m,$ that 
$
\|\hat{\bbeta}_j\hat{\bbeta}_j^T-\bbeta_j\bbeta_j^T\|_F^2=2\{1+o_P(1)\}\delta \lambda_j+\delta^2+O_P(\lambda_j^2n^{-1}).
$
Here, from (\ref{D**}) and (\ref{D}), 
it holds that $\hat{\lambda}_{j}^{1/2}\hat{\bh}_{j}^T \bh_{j'}=\lambda_{j'}^{1/2} \bu_{j'}^T\hat{\bu}_{j}=O_P(\lambda_{j'}^{1/2}/n^{1/2})$. 
Thus, we can conclude the result in Proposition A.2. 
\end{proof}
\CG{
\begin{proof}[Proof of Proposition B.1]
Assume (A-i),  
(C-i), (C-ii) and ($\star$'). 
By using Markov's inequality, 
for any $c>0$, 
under (A-ii'), we have as $d\to \infty$ and $n\to \infty$ that  
\begin{align*}
\sum_{j'=1}^dP(|\bar{z}_{j(j'),x}|\ge  cd^{1/8}/n^{1/2} )
&=
\sum_{j'=1}^dP(|\bar{z}_{j(j'),x}|^8\ge  c^8d/n^{4} )  \notag  \\
& \le \sum_{j'=1}^d \frac{n^4E(|\bar{z}_{j(j'),x}|^8)}{d c^8} 
=O(c^{-8}) \ \mbox{ for $j=1,...,m$; \ and} \\
\sum_{j'=1}^dP( \|\bx_{(j'),2} \| \ge  c d^{1/8} )&=
\sum_{j'=1}^dP( \|\bx_{(j'),2} \|^8 \ge  c^8 d ) \notag  \\
&\le \sum_{j'=1}^d\frac{E(\|\bx_{(j'),2} \|^8)}{d c^8}=O(c^{-8}),
\end{align*}
so that  
\begin{align*}
\bar{z}_{j(j'),x}=O_P(  d^{1/8}/n^{1/2}) \ \mbox{ for all $j,j'$; \ and} \notag  \\
\|\bx_{(j'),2} \| =O_P (  d^{1/8}) \ \mbox{ for all $j'$}.
\end{align*}
Then, from (\ref{D*}) and (\ref{D7}), 
it holds for $j=1,...,m$, that 
\begin{equation}
\tilde{\lambda}_{j}^{-1/2}\bx_{(j'),2}^T \hat{\bu}_{j}= O_P\{( {\lambda}_{j}^{-1}  n^{-1}d^{1/4})^{1/2}\}  
\  \mbox{ for all $j'$.} \notag 
\end{equation}
Thus from (\ref{D}), 
by combining (\ref{D8}) and (\ref{D9}), 
we have for $j=1,...,m$, that
\begin{equation}
\tilde{h}_{j(j')}={h}_{j(j')}+O_P\{({\lambda}_{j}^{-1}  n^{-1}d^{1/4})^{1/2}  \}  \  \mbox{ for all $j'$.} 
\label{DDD1}
\end{equation}
From (\ref{DDD1}), 
under (C-iii), 
it holds for $j=1,...,m$ and all $j'$, that 
\begin{align}
\tilde{h}_{j(j')}^2={h}_{j(j')}^2+o_P\Big({h}_{oj(k_{j*})} \max\{ |{h}_{oj(k_{j*})}|,|{h}_{j(j')}| \}\Big). 
\label{DDD2}
\end{align}
From  (2.3), note that $\lambda_j^{-1}d^{1/4}=o(1)$ for $j=1,...,m$. 
Then, from (\ref{DDD2}), similar to the proofs of Theorem 1 and Corollary 1, 
we can obtain the results in Theorem 1 and Corollary 1 after replacing (A-ii) and ($\star$) with (A-ii') and ($\star $'). 
By using the results in Theorem 1 and Corollary 1,  similar to the proofs of Theorem A.1 and Corollary A.1, 
we can obtain the results in Theorem A.1 and Corollary A.1 after replacing (A-ii) and ($\star$) with (A-ii') and ($\star $'). 
For Theorem 2, from (\ref{DDD1}), similar to the proofs of Theorem 2, 
we can obtain the result after replacing (A-ii) and ($\star$) with (A-ii') and ($\star $'). 
\end{proof}
}

\end{document}